\documentclass[11pt,a4paper]{article}
\usepackage{jheppub}
\usepackage{slashed}
\usepackage{amsmath}
\usepackage{amssymb}
\usepackage{epsfig}
\usepackage{subfig}

\newcommand{\labell} [1] {\label{#1}}

\def\({\left(} 
\def\){\right)}
\def\[{\left[} 
\def\]{\right]}
\newcommand{\non}{\nonumber \\}

\newcommand{\ie}{{\it i.e.,}\ }
\newcommand{\eg}{{\it e.g.,}\ }

\newcommand{\A}{\mathcal{A}}

\newcommand{\cR}{{\mathcal R}}

\newcommand{\E}{\mathcal{E}}
\newcommand{\cM}{{\mathcal M}}

\newcommand{\C}{{\mathcal C}}

\newcommand{\X}{\mathcal{X}}
\newcommand{\B}{\mathcal{B}}

\newcommand{\cL}{\mathcal{L}}

\newcommand{\ren}{R\'enyi\ }

\newcommand{\be}{\begin{equation}}
\newcommand{\ee}{\end{equation}}
\newcommand{\bea}{\begin{eqnarray}}
\newcommand{\eea}{\end{eqnarray}}

\newcommand{\mt}[1]{\textrm{\tiny #1}}

\newcommand{\tr}{{\rm tr}}

\newcommand{\lp}{\ell_{\mt P}}

\newcommand{\fin}{f_\infty}

\catcode`\@=12

\newcommand{\al}{\alpha}

\def\del          {\partial}

\def\tr           {\mathop{\rm Tr}}

\newcommand{\reef}[1]{(\ref{#1})}
\renewcommand{\eqref}[1]{(\ref{#1})}

\def\ph1{\phantom{1}}

\newcommand{\beq}{\begin{equation}}
\newcommand{\eeq}{\end{equation}}
\newcommand{\ba}{\begin{aligned}}
\newcommand{\ea}{\end{aligned}}
\newcommand{\beqa}{\begin{eqnarray}}
\newcommand{\eeqa}{\end{eqnarray}}
\newcommand{\beqar}{\begin{eqnarray*}}
\newcommand{\eeqar}{\end{eqnarray*}}

\newcommand\te{t_\mt{E}}
\newcommand{\rhoa}{\rho_\mt{A}}

\def\Tr{\mathop{\mathrm{Tr}}}

\newcommand{\taue}{\tau_\mt{E}}
\newcommand{\tsigma}{\tilde{\sigma}}

\newcommand{\lgb}{\lambda_\mt{GB}}
\newcommand{\mqt}{\mu_\mt{QT}}

\newcommand{\aha}{\hat{a}}
\newcommand{\bha}{\hat{b}}
\newcommand{\cha}{\hat{c}}

\newcommand{\Hm}{H_m}

\arxivnumber{1407.xxxx}

\title{Twist operators in higher dimensions}

\author[a,d]{Ling-Yan Hung,}
\author[b]{Robert C. Myers}
\author[c]{and Michael Smolkin}

\affiliation[a]{Department of Physics, Harvard University, Cambridge MA 02138, USA}
\affiliation[b]{Perimeter Institute for Theoretical Physics, Waterloo, Ontario N2L 2Y5, Canada}
\affiliation[c]{Center for Theoretical Physics and Department of Physics,
University of California, Berkeley, CA 94720, USA}
\affiliation[d]{Department of Physics, Fudan University, Shanghai, China}

\emailAdd{lhung@physics.harvard.edu}
\emailAdd{rmyers@perimeterinstitute.ca}
\emailAdd{smolkinm@berkeley.edu}

\abstract{We study twist operators in higher dimensional CFT's. In particular, we express their conformal dimension in terms of the energy density
for the CFT in a particular thermal ensemble. We construct an expansion of the conformal dimension in power series around $n=1$,
with $n$ being replica parameter. We show that the coefficients in this expansion are determined by higher point correlations of the energy-momentum tensor.
In particular, the first and second terms, \ie the first and second derivatives of the scaling dimension, have a simple universal form.  
We test these results using holography and free field theory computations, finding agreement in both cases. We also consider the `operator product expansion' 
of spherical twist operators and finally, we examine the behaviour of
correlators of twist operators with other operators in the limit $n\to1$.}

\begin{document}

\maketitle

\section{Introduction}

Recently, entanglement entropy and related theoretical tools have received considerable attention in areas ranging from condensed matter physics,
\eg \cite{wenx,cardy0} to quantum gravity, \eg \cite{mvr,arch,holes}. In particular, holographic
entanglement entropy \cite{rt0} is now playing an important
role in developing our understanding of gauge/gravity duality. This concept has evolved to become a universal tool that intertwines 
the non-perturbative structure of the boundary field theory and the quantum nature of the bulk spacetime. 

A challenge to gaining a better understanding of entanglement entropy in quantum field theory (QFT) remains simply a deficiency of computational tools. 
One of the most commonly used techniques for evaluating entanglement entropy is the `replica trick' \cite{cardy0}. In this approach, given the reduced
density matrix $\rhoa$ describing the QFT restricted to a certain spatial region $A$, one first evaluates the \ren entropies
  \be
S_n={1\over 1-n}\,\log\, \Tr \rhoa^{\,n}= {1\over n-1}\,\(n\,\log Z_1 - \log Z_n\)\,.
  \labell{ren2}
  \ee
where $Z_n$ is the partition function on an $n$-fold covering geometry, with cuts introduced on region $A$. The entanglement entropy 
is then determined as the limit: $S_\mt{EE}=\lim_{n\to1} S_n$. In this construction, the entangling surface
$\Sigma$, which encloses the region $A$, becomes the branch-point of the cut which separates different copies
in the $n$-fold cover. It is convenient to think of these
boundary conditions as produced by the insertion of a ($d$--2)-dimensional
surface operator, \ie the twist operator $\sigma_n$, at $\Sigma$  which interlaces $n$ copies
of the QFT on a single copy of the background geometry \cite{cardy0}. These twist operators will be in the focus of our current study.

In two (spacetime) dimensions, twist operators are local operators \cite{cardy0} and for a two-dimensional conformal
field theory (CFT), they are in fact conformal primaries whose scaling dimension is given by
 \be h_n=
 \frac{c}{12} \(n -\frac{1}{n}\)\,.
 \labell{dim}
 \ee
However, in general dimensions, the replica-trick construction provides only a formal definition of twist
operators for higher dimensions and so in practice beyond $d=2$, the
properties of these operators is not well understood. 
In \cite{renyi}, a holographic study suggested a new approach to evaluate a generalized notion of
the conformal dimension $h_n$ of the twist operators in higher dimensional CFT's, which we will
review below. Further, this work \cite{renyi} revealed in a wide variety of holographic theories,
this conformal weight satisfied a simple and intriguing relation:
 \be
\partial_n h_n|_{n=1} = 2 \pi^{\frac{d}{2}+1}\,
 \frac{\Gamma \( {d}/{2}\)}{\Gamma(d+2)}\ C_T\,,
 \labell{delhx}
 \ee
where $C_T$ is the central charge appearing in the two-point function of the stress tensor --- see eq.~\reef{emt2p} below.
One of our results here is to explain that this expression is, in fact, universal 
applying for twist operators in any CFT.  Further, we will also show that
$\del^2_n h_n|_{n=1}$ has a similar universal form involving the CFT parameters
which determine the three-point function of the stress tensor.

The remainder of the paper is organized as follows:
In section \ref{twist}, we review the construction of \cite{renyi,circle4} which allows us to evaluate the
scaling dimension of twist operators in higher dimensional CFT's in terms of the energy density of a thermal ensemble
on a certain hyperbolic geometry. Next, we use the resulting expression to make an expansion of the conformal dimension in power series around $n=1$,
and show that the $k$-th order coefficient is detemined by the $(k+1)$- and $k$-point correlation functions of the stress tensor.
As noted above, this yields simple universal expressions for the first and second terms, \ie the first and second derivatives of the scaling dimension.
In section \ref{eric}, we extend these results to directly expand the \ren entropy about $n=1$, as was considered recently in \cite{Perlmutter0}.
We find our results are in complete agreement with the latter reference.  In section \ref{compare}, we compare our results for $\del_n h_n|_{n=1}$ and $\del^2_n h_n|_{n=1}$
with explicit computations in various holographic models and in free field theories. Section \ref{OPE}
reviews the generalized OPE expansion for twist operators and we extract certain coefficients in the OPE expansion of a spherical twist operator.
In section \ref{small}, we propose that the twist operator can be effectively represented by a construction involving the modular Hamiltonian.
This construction allows us to consider an expansion for small $(n-1)$ of the twist operators themselves. 
Finally, we close with a brief discussion of our results and future
directions in section \ref{discuss}.  The evaluation of a useful integral used in section \ref{twist} is presented in appendix \ref{integralx}. 
Further the details of heat kernel computations of the conformal dimensions of twist operators in free theories are presented in appendix \ref{appfree}.

\section{Twist operators for higher dimensional CFT's} \labell{twist}

As discussed above, just as in two dimensions,
twist operators are naturally defined in higher dimensions through the replica 
trick. In $d$ dimensions then, 
a twist operator $\sigma_n$ is a ($d-2$)-dimensional \emph{surface}
operator which introduces a branch cut at the entangling surface in the path integral over
the $n$-fold replicated theory. In this section, we examine this formal definition 
more carefully to produce certain explicit results for these surface operators. However, we
will restrict our attention to twist operators in higher dimensional CFT's.
More specifically, we consider a CFT in its vacuum state in $d$-dimensional flat space
and we choose the entangling surface to be a sphere of radius $R$ (on a constant time slice).
In this case, it was shown that the entanglement entropy can be evaluated
as the thermal entropy of the CFT on a hyperbolic cylinder $R\times H^{d-1}$, where the
temperature and curvature are fixed by the radius of the original entangling
surface \cite{circle4}. Further, this approach is easily extended to a calculation of the \ren
entropies by varying the temperature in this thermal ensemble \cite{renyi}. As will be
described in the following, we use this construction here to produce a better understanding
of the corresponding twist operators.

\subsection{Spherical twist operators} \labell{sphere}

A key point in the analysis of \cite{renyi,circle4} is to take advantage of the fact
that the underlying theory is a conformal field theory and to find a conformal transformation
mapping the theory between flat space and the hyperbolic cylinder. Hence, we begin with a review
of this transformation for the corresponding Euclidean signature geometries: Following \cite{renyi},
the metric on flat (Euclidean) space may be written in terms of a  complex coordinate 
$\omega=r+i \te$:
 \be
ds^2_{R^d} = d\te^2 + dr^2 +r^2d\Omega_{d-2}^2
= d\omega d\bar{\omega} + \left(\frac{\omega +\bar{\omega}}{2}
\right)^2 d\Omega_{d-2}^2\,,
 \labell{flat0}
 \ee
where $d\Omega_{d-2}^2$ denotes a standard round metric on a
unit $(d-2)$-sphere. The spherical entangling surface will be located  at
$(\te,r)=(0,R)$, or in terms of the complex coordinate, at $\omega=R$.
Now to construct the desired conformal
transformation, we introduce a second complex coordinate $\sigma = u + i \frac{\taue}{R}$ 
and then make the coordinate transformation
 \be
e^{-\sigma} = \frac{R-\omega}{R+\omega} \,. \labell{change}
 \ee
The metric \reef{flat0} then becomes
 \be
ds^2_{R^d}= \Omega^{-2}\,R^2\left[d\sigma d\bar{\sigma} +
\sinh^2\left(\frac{\sigma + \bar{\sigma}}{2}\right) d\Omega_{d-2}^2\right]\,,
 \labell{golf}
 \ee
where
 \be
\Omega = \frac{2R^2}{|R^2-\omega^2|}=|1+\cosh\sigma|\, .
 \labell{factx}
 \ee
Removing the $\Omega^{-2}$ prefactor by a simple Weyl rescaling,
the resulting metric reduces to
 \be
ds^2_{H^{d-1}\times S^1} = \Omega^2\, ds^2_{R^d} = d\taue^2+R^2\left(du^2 +
\sinh^2\! u\, d\Omega_{d-2}^2\right)\,. \labell{frog}
 \ee
This conformally transformed  geometry corresponds to  $S^1\times H^{d-1}$,
where $u$ is the (dimensionless) radial coordinate on the hyperboloid
$H^{d-1}$ and $\taue$ is the Euclidean time coordinate on $S^1$. As is clear from eq.~\reef{frog},
the curvature radius of $H^{d-1}$ is $R$, the radius of the original spherical entangling surface.

To confirm that the $\taue$ direction is indeed periodic, we can examine the transformation \reef{change}
in the vicinity of the entangling surface. That is, if we choose $\omega=R-\delta r - i\delta\te$ with 
$\delta r, \delta\te\ll R$, then to leading order, eq.~\reef{change} becomes
 \be
e^{-u-i \frac{\taue}{R}} \simeq \frac{\delta r + i\delta\te}{2R} \,. \labell{change2}
 \ee
This expression is the usual exponential mapping between the plane $R^2$ and the
cylinder $R\times S^1$. Hence it makes clear that the $\taue$ coordinate lives on a circle 
and further that we should identify the period as $\Delta\taue=2\pi R$ to ensure that the
geometry is smooth at the entangling surface, \ie at the origin of the ($\delta r,\delta\te$)-plane. 
We also note that from eq.~\reef{change2}, it is apparent
that the entangling surface at $\omega=R$ has
been pushed out to $u\to\infty$, the asymptotic boundary of the hyperbolic geometry.
 
Given the periodicity of the Euclidean time coordinate $\taue$, it is evident that the CFT 
on the  hyperbolic geometry is at finite temperature with
 \beq
 T_0=\frac{1}{2\pi R}.
 \labell{bus}
 \eeq
Hence under the conformal mapping, the reduced density matrix
describing the CFT on the interior of the spherical
entangling surface is transformed to a thermal
density matrix,
 \beq
\rhoa=U^{-1}\,\rho_{\mt{thermal}}(T_0)\,U=U^{-1}\, \frac{e^{-H/T_0}}{Z(T_0)}\,U \,. \labell{triangle}
 \eeq
Here $U$ denotes the unitary transformation implementing the conformal
transformation. Since the entropy is insensitive to such a unitary transformation,
the desired entanglement entropy just equals the thermal
entropy in the transformed space \cite{circle4}. 

Now to evaluate the \ren entropy $S_n$ as in eq.~\reef{ren2}, we must consider the $n$'th power of
the density matrix
 \beq
\rhoa^{\,n}=U^{-1}\, \frac{e^{-n\,H/T_0}}{Z(T_0)^n}\,U \,. \labell{trianglen}
 \eeq
That is, we must consider the thermal ensemble with the temperature $T=T_0/n$
on the same hyperbolic geometry. Hence the period $\Delta\taue$ is extended to $2\pi n\,R$.
Now applying the same conformal mapping \reef{change} in this case, will again yield the
flat space metric \reef{flat0}. However, examining the geometry near the entangling surface
with eq.~\reef{change2}, we see that the origin is now circled $n$ times as $\taue$ runs over
its full period. Therefore,
as might have been anticipated, the transformation
\reef{change} actually maps the thermal background $S^1\times H^{d-1}$
to an $n$-fold cover of $R^d$ with an orbifold singularity located precisely
at the entangling surface, \ie the ($d-2$)-dimensional sphere given by $r=R$
(and $\te=0$). Hence the path integral on this new geometry would yield precisely the partition
function $Z_n$ of an $n$-fold replicated theory with a spherical twist
operator inserted at $r=R$. 

While this twist operator is the focus of our study, let us add that 
since we are studying a CFT, we may equate 
$Z_n=Z(T_0/n)$ because the two path integrals are simply related by a conformal 
transformation.\footnote{Of course,
with the choice $n=1$, we have the original one-to-one mapping from
$S^1\times H^{d-1}$ to $R^d$.}
Hence the \ren entropy \reef{ren2} may be re-expressed in terms of these thermal 
partition functions,
  \be
S_n= {1\over n-1}\,\(n\,\log Z(T_0) - \log Z(T_0/n)\)\,.
  \labell{ren3}
  \ee
Then using the standard
thermodynamic identity $S_{\text{therm}}(T)=\partial_T\[T\log Z(T)\]$, we
may express the \ren entropy in terms of the thermal entropy \cite{renyi},
 \beq
S_{n}\ =\ {n\over n-1}{1\over
T_{0}}\int_{T_{0}/{n}}^{T_{0}}S_{\text{therm}}(T)\ dT\,.
 \labell{Rnyi-therm}
 \eeq

\subsection{Conformal dimension} \labell{dimens}

As noted in the introduction, in a higher dimensional CFT, the twist operators
may be assigned a generalized notion of conformal dimension \cite{anton}. As in $d=2$,
the latter is defined by the leading singularity in the correlator 
$\langle T_{\mu\nu}\, \sigma_n \rangle$. We review the structure of this singularity here
following the discussion in \cite{renyi}: First in flat (Euclidean) space,
we make an insertion of the stress tensor in the vicinity of a {\it planar} twist operator $\sigma_n$.
We align the Cartesian coordinates $x^\mu$ on $R^d$ with 
the twist operator, so that this surface operator is positioned at $x^1=0=x^2$ 
while it extends throughout the remaining coordinates with $\mu=a\in\{3,\ldots,d\}$. 
With the stress tensor inserted at $x^\mu=\{y^i,x^a\}$, the perpendicular distance
to the twist operator is defined as $y=\sqrt{(y^1)^2+(y^2)^2}$. Now symmetry
dictates the form of the corresponding correlator up to a single constant, \ie
the conformal dimension. Specifically,  the basic geometric
structures appearing in the correlator are determined by the residual translational and rotational
symmetries, which remain in the presence of the twist operator. Then the relative normalization of various
contributions is fixed by the tracelessness and
conservation of the stress tensor, \ie by imposing $\langle T^\mu{}_{\mu}\, \sigma_n
\rangle=0=\nabla^\mu \langle T_{\mu\nu}\, \sigma_n \rangle$. 
Subject to all of these constraints, the correlator is restricted to take the following 
form\footnote{These expressions are implicitly
normalized by dividing by $\langle \sigma_n \rangle$ but we left
this normalization implicit to avoid the clutter that would otherwise be created.}
 \bea
\langle T_{ab}\, \sigma_n \rangle &=& -\frac{h_n}{2\pi}\,
\frac{\delta_{ab}}{y^d}\,, \qquad \langle T_{ai}\, \sigma_n\rangle = 0\,,
\labell{weight}\\
\langle T_{ij}\, \sigma_n\rangle &=& \frac{h_n}{2\pi}\, \frac{(d-1)\delta_{ij} -
d n_i n_j}{y^d}\,,
 \nonumber
 \eea
where $a,b$ ($i,j$) denote tangential (normal) directions to the twist operator
and $n^i=y^i/y$ is the unit vector directed orthogonally from the twist operator to
the $T_{\mu\nu}$ insertion. Thus the correlator is completely fixed up to the single
constant  $h_n$. The latter is commonly referred to as the conformal dimension of
$\sigma_n$, since its
appearance above is analogous to that of the scaling dimension of a
local primary operator. In particular then,
if one reduces these expressions to $d=2$ (in which case the twist operators
are local primaries), one finds that the present
definition for $h_n$ matches with the standard definition, as given in \cite{cardy0},
and $h_n$ is precisely the total scaling dimension given in eq.~\reef{dim}.
Further, note that we are assuming that $T_{\mu\nu}$ corresponds
to the total stress tensor for the entire $n$-fold replicated CFT, \ie
$T_{\mu\nu}$ is inserted on all $n$ sheets of the universal cover. 

Given the basic definition of $h_n$, we can now use the conformal mapping
described in the previous section to gain further insights about this parameter \cite{renyi}. On one
side of this mapping, we have the CFT in a thermal ensemble on the hyperbolic
cylinder or rather the Euclidean CFT lives on the background
$S^1\times H^{d-1}$. On general grounds, the
expectation value of the stress tensor will then take the form
 \begin{equation}
 \langle T^\mu{}_\nu\rangle=\textrm{diag}(-\E(T),\,p(T),\,\cdots,\,p(T))\,,
 \labell{none5}
 \end{equation}
where the energy density $\E(T)$ and the pressure $p(T)$ 
are constant throughout the hyperbolic background.\footnote{The 
signs are chosen here in eq.~\reef{none5} for a Euclidean signature,
\ie $ \mathcal{E}(T)=-\langle T_{\taue \taue}\rangle$.} 
Further the trace of this expression must vanish in a CFT\footnote{In principle,
the trace anomaly could lead to a nonvanishing trace in even dimensions.
However, one can readily verify that in fact the
trace anomaly vanishes for the background geometry $S^1\times H^{d-1}$.
In particular, the Euler density vanishes because the
background is the direct product of two lower dimensional geometries.
Further this background is conformally flat and so any conformal
invariants also vanish.} and hence,
\begin{equation}
p(T)=\E(T)/(d-1)\,. \labell{none6}
\end{equation}

Next we can relate the thermal energy density to the correlator \reef{weight}
by applying the conformal map from $S^1\times H^{d-1}$ to the $n$-fold
cover of $R^d$ described above. In particular, recall that the $n$-fold cover is
produced when the temperature is tuned to $T=T_0/n$. Now under this 
conformal mapping, the stress tensor becomes
 \be
\langle T_{\al\beta}\ \sigma_n \rangle=\Omega^{d-2}\,\frac{\partial X^\mu}{\partial
x^\al}\,\frac{\partial X^\nu}{\partial x^\beta}\,\Big(
\langle T_{\mu\nu}(T_0/n)\rangle
- {\A}_{\mu\nu}\Big)\, ,
 \labell{stressd}
 \ee
where $\alpha,\beta$  and $\mu,\nu$ denote indices on the flat geometry
and $S^1\times H^{d-1}$, respectively. Since the conformal mapping
generates an orbifold singularity in the (otherwise) flat covering space,
the expectation on the left-hand side
of eq.~\reef{stressd} has been interpreted as the expectation value
of the stress tensor in the presence of the corresponding
twist operator $\sigma_n$ (on the sphere at $r=R$ and $\te=0$).
Further, the stress tensor is not a primary operator and so an anomalous 
contribution $\A_{\mu\nu}$ also appears on the right-hand side.
This contribution is the higher dimensional analog of the usual Schwarzian term
appearing in two dimensions \cite{skenny}. We observe that this anomalous term
depends entirely on the details of the transformation \reef{change}, but it is
independent of the temperature in the hyperbolic background, \ie the period of
the $S^1$. Therefore this term can be fixed by noting that eq.~\reef{change} produces a
one-to-one mapping from $S^1\times H^{d-1}$ to $R^d$ with $n=1$. Hence
since there is no orbifold singularity in this case, the left-hand side becomes
$\langle T_{\al\beta}\rangle$, \ie the vacuum expectation value of the stress
energy in flat space, and so it simply vanishes. Since the left-hand side vanishes with 
$n=1$, we conclude that ${\A}_{\mu\nu}=T_{\mu\nu}(T_0)$. Hence eq.~\reef{stressd} 
becomes
 \be
\langle T_{\al\beta}\ \sigma_n \rangle =\Omega^{d-2}\,\frac{\partial X^\mu}{\partial
x^\al}\,\frac{\partial X^\nu}{\partial x^\beta}\,\Big(\langle T_{\mu\nu}(T_0/n)\rangle
-\langle T_{\mu\nu}(T_0)\rangle\Big)\, .
 \labell{stressfin}
 \ee
Recall that the conformal factor $\Omega$ is given by eq.~\reef{factx}.

Note that the conformal mapping above generates a {\it spherical} twist
operator while conformal dimension was defined in eq.~\reef{weight}
by the correlator of the stress tensor with a {\it planar} twist operator. However, $h_n$
can be identified here by bringing the insertion of the stress tensor 
very close to the spherical twist operator, in which case
the leading singularity in eq.~\reef{stressfin} will emerge 
with the same form as in eq.~\reef{weight}. To evaluate this
singularity, we begin by examining eq.~\reef{change} which yields
\bea
\frac{\partial u}{\partial \te}&=&-{1\over R}\frac{\partial \taue}{\partial r}={iR(\omega^2-\bar\omega^2)\over (R^2-\omega^2)(R^2-\bar\omega^2)}~ , \non
\frac{\partial u}{\partial r}&=&\ \ {1\over R}\frac{\partial \taue}{\partial \te}= {2R^3-R(\omega^2+\bar\omega^2)\over (R^2-\omega^2)(R^2-\bar\omega^2)}~.
\labell{CReq}
\eea
Of course, the first equality in each of the above expressions 
corresponds to the standard Cauchy-Riemann conditions.
Next, to simplify the analysis, we insert $T_{\al\beta}$ at $t_\mt{E}=0$
and $r = R-y$ with $y \ll R$ (as well as some fixed angles).
With this choice, eq.~\reef{CReq} simplifies to $\frac{\partial
u}{\partial \te} = 0$ and $\frac{\partial
\taue}{\partial \te}\simeq R/y$ and 
further, eq.~\reef{factx} gives $\Omega\simeq R/y$. 
Given these expressions and setting $\al=t_\mt{E}=\beta$, eq.~\reef{stressfin} yields
 \bea
\langle T_{t_\mt{E}t_\mt{E}}\, \sigma_n
\rangle &=&\Omega^{d-2}\,\(\frac{\partial
\tau_\mt{E}}{\partial
t_\mt{E}}\)^2\,\Big(T_{\tau_\mt{E}\tau_\mt{E}}(T_0/n)
-T_{\tau_\mt{E}\tau_\mt{E}}(T_0)\Big)
 \nonumber\\
&=& -\(\frac{R}{y}\)^d\,\Big(\E(T_0/n) -\E(T_0)\Big) +\cdots\,.
 \labell{last}
 \eea
This result should be compared to the $i=j=\te$ component in
eq.~\reef{weight}, \ie 
\be
\langle T_{t_\mt{E}t_\mt{E}}\, \sigma_n\rangle= \frac{d-1}{2\pi}\ \frac{h_n}{y^d}\,.
\labell{compareX}
\ee
However, first, we recall that the 
expectation value in eq.~\reef{compareX} involves the total stress tensor for the entire
$n$-fold replicated CFT, while in eq.~\reef{last},
we have an insertion of $T_{\te\te}$ on a single sheet of the
universal cover. Hence the latter must be multiplied by an extra factor of $n$ before
comparing the two expressions. The final result for the scaling dimension is
 \be
h_n=\frac{2\pi\,n}{d-1}\,R^d\,\Big(\E(T_0)-\E(T_0/n) \Big)\,.
 \labell{heavy}
 \ee

We now turn to the intriguing result \reef{delhx} which was found in \cite{renyi} to apply for 
a variety of holographic models:
 \be
 \partial_n h_n|_{n=1} = 2 \pi^{\frac{d}{2}+1}\,
 \frac{\Gamma \( {d}/{2}\)}{\Gamma(d+2)}\ C_T\,.
 \labell{interest1}
 \ee
Here $C_{T}$ is the central charge defined by the two-point function of the
stress tensor.\footnote{Note that our normalization for $C_T$ above corresponds to that
introduced in \cite{Osborn0,Erdmenger0} but not the same as in \cite{renyi}. Hence the numerical
factor in eq.~\reef{interest1} is  different than given in the original
reference \cite{renyi}.} In fact, we will now show below that eq.~\reef{interest1}
is a universal result that applies for the scaling dimension of twist operators in any CFT.

Our proof that eq.~\reef{interest1} is universal begins with eq.~\reef{heavy}, which again
applies for any CFT. The energy densities in the latter equation are evaluated
in a thermal ensemble on the hyperbolic cylinder and so can be written as
 \beq
\E(T)=-\langle T_{\taue\taue}\rangle=-\Tr\[\rho_{\mt{thermal}}(T)\ T_{\taue\taue}\]=
 -\frac{1}{Z(T)}\,\Tr\[ e^{-H/T}\,T_{\taue\taue}\] \,. \labell{triaX}
 \eeq
Combining eqs.~\reef{heavy} and \reef{triaX}, the expression on the left-hand side of
eq.~\reef{interest1} becomes 
 \beqa
 \del_n h_n|_{n=1}&=&-{2\pi R^d \over (d-1)T_0}  \langle H\  T_{\taue \taue}(x_0)\rangle_c 
 \labell{Xder}\\
&=&{2\pi R^d \over (d-1)T_0} \ \int_{H^{d-1}}\!\!\!d^{d-1}\tilde{x}\,\sqrt{\gamma}\ \langle T_{\taue \taue}(\tilde{x})
\  T_{\taue \taue}(x_0)\rangle_c
\nonumber
 \eeqa
where $\gamma$ stands for the determinant of the induced metric on the Cauchy surface on which $H$ is
evaluated and the subscript `$c$' denotes the connected part of the thermal correlator, \ie
 \beq
\langle T_{\taue \taue}(\tilde{x})
\  T_{\taue \taue}(x_0)\rangle_c =\langle T_{\taue \taue}(\tilde{x})
\  T_{\taue \taue}(x_0)\rangle-\langle T_{\taue \taue}(\tilde{x})\rangle\ 
\langle  T_{\taue \taue}(x_0)\rangle \,.
 \labell{connect}
 \eeq
Further with $n$ set to 1 on the left-hand side of eq.~\reef{Xder}, we must evaluate the corresponding
thermal expectation values at $T=T_0$. 

Now, given the expression in eq.~\reef{Xder} where $\del_n h_n|_{n=1}$ is expressed
in terms of a two-point function of the stress tensor, it is natural to expect that the final result should
be proportional to $C_T$. However, we may further evaluate the precise constant of proportionality in
order to establish the universality of the result in eq.~\reef{interest1}.

In eq.~\reef{Xder}, we introduced a specific location $x^\al_0$ for the insertion of the stress tensor. Of
course, the choice of this location is arbitrary since the thermal bath on the hyperbolic geometry is homogeneous.
Similarly, in the final expression, the Hamiltonian can be evaluated with an integration over any Cauchy
surface because the stress tensor is conserved. To simplify the following analysis, we fix the location $x_0^\al$ to be
at $u=0$ and $\taue=0$ and also evaluate the Hamiltonian by integrating over the surface $\taue=0$. With
this choice, the correlator in eq.~\reef{Xder} is spherically symmetric and so using eq.~\reef{frog}, we may write
 \beq
 \del_n h_n|_{n=1}={2\pi R^{2d-1} \over (d-1)T_0} \ \Omega_{d-2}\int du\ \sinh^{d-2}\!u\ \langle T_{\taue \taue}(\taue=0,u)
\  T_{\taue \taue}(\taue=0,u=0)\rangle_c 
\labell{Xder2}
 \eeq
where $\Omega_{d-2}$ denotes the volume of a unit
($d$--2)-sphere.

Now using the conformal transformation described in the previous two sections, we can map the two-point
correlator of the stress tensor on the Euclidean background $S^1\times H^{d-1}$ with $\beta=1/T_0$
to the two-point correlator in the CFT vacuum on $R^d$. That is, we can relate the thermal correlator in
eq.~\reef{Xder} to the usual two-point correlator in flat space which defines $C_T$ --- see below. In particular, with
special choice of insertion points described above, eq.~\reef{stressfin} 
yields\footnote{Note that the anomalous terms do not contribute here in the transformation
of the connected correlator.}
 \beqa
 \langle T_{\taue \taue}(0,u)\, T_{\taue \taue}(0,0)  \rangle_c
 &=&\left[\frac1{\Omega^{d-2} }\(\frac{\del\te}{\del\taue}\)^2\right]_{\taue=0,u}\ 
\left[\frac1{\Omega^{d-2} }\(\frac{\del\te}{\del\taue}\)^2\right]_{\taue=0,u=0}
 \labell{2ptensor}\\
&&\qquad\qquad
\times
\ \langle  T_{\te\te}(\te=0,r) \ T_{\te\te}(\te=0,r=0)  \rangle_c\Big|_{r=R\tanh u/2} ~,
 \nonumber
 \eeqa
where eq.~\reef{change} was used to determine $(\taue=0,u)\to(\te=0,r=R\tanh u/2)$.
Implicitly, we have also used eq.~\reef{CReq} to show $\left.\del r/\del\taue\right|_{\taue=0}=0$.
Further, this equation also yields
 \be
\frac{\del\te}{\del\taue}\Big|_{\taue=0}=\frac{R^2-r^2}{2\,R^2}=\frac1\Omega\Big|_{\te=0,r\le R}~.
\labell{very}
 \ee
Note that we have the following additional simplifications for $\te=0$:  $\sinh u=\Omega\, r/R$ and
$\del u/\del r =\Omega/R$, as well as $\Omega|_{\te=0,r=0}=2$.
Combining these results, eq.~\reef{Xder2} becomes
 \beq
 \del_n h_n|_{n=1}={ R^{d-1} \over (d-1)2^{d}T_0} \ \Omega_{d-2}\,2\pi\int_0^R
 dr\,r^{d-2}\,\frac{R^2-r^2}{2R} 
\langle T_{\te \te}(\te=0,r)
\  T_{\te \te}(\te=0,r=0)\rangle_c \,.
\labell{Xder3}
 \eeq
Since this correlator is now evaluated in $R^d$, we may drop the subscript $c$  since the corresponding
one-point expectation values vanish --- in particular, $\langle T_{\te \te}\rangle=0$.
Further, we observe that the standard Hamiltonian on the hyperbolic space appearing
in eq.~\reef{Xder} has been transformed in the above expression
to the corresponding entanglement Hamiltonian for a spherical region in flat space found in 
\cite{circle4}\footnote{More precisely, the conformal mapping takes $H/T_0\to \Hm$ \cite{gogo} --- see eq.~\reef{modu} below.}

Now recall that the two-point function of the stress tensor for a general CFT in $R^d$ takes the form \cite{Osborn0,Erdmenger0} 
 \be
  \langle  T_{\mu\nu}(x)\,  T_{\alpha\beta}(0)  \rangle = {C_T \over x^{2d}}\, 
\mathcal{I}_{\mu\nu,\alpha\beta}(x)~,
  \labell{emt2p}
 \ee
where $C_T$ is a constant and the tensor structure is given by
 \be
 \mathcal{I}_{\mu\nu,\alpha\beta}={1\over 2}(I_{\mu\alpha}\,I_{\nu\beta}+I_{\mu\beta}\,I_{\nu\alpha})
-{1\over d} \delta_{\mu\nu}\,\delta_{\alpha\beta}\qquad
{\rm with}\ \ 
 I_{\mu\nu}(x)=\delta_{\mu\nu}-2{x_{\mu}x_{\nu}\over x^2}~.
\labell{dunce1}
 \ee
Hence the desired correlator becomes
 \be
 \langle T_{\te\te}(0,r)\, T_{\te\te}(0,0)  \rangle={d-1\over d}\,{C_T \over r^{2d}}~.
\labell{dunce2}
 \ee
Now substituting this expression, as well as $\Omega_{d-2}=2\pi^{(d-1)/2}/\Gamma\left({(d-1)/ 2}\right)$ and 
$T_0=1/(2\pi R)$, into eq.~\reef{Xder3}, we find the final result
  \beqa
 \del_n h_n|_{n=1}&=&{\pi^2 R^{d+1} \over d\,2^{d-2}} \ \frac{2\pi^{(d-1)/2}}{\Gamma\left({(d-1)/ 2}\right)}
\int_0^R dr\,r^{d-2}\,\frac{R^2-r^2}{2R^2} 
\,{C_T \over r^{2d}}\nonumber\\
&=& \frac{\pi^{(d+3)/2}}{2^{d-3}\,d
\,(d^2-1)\,\Gamma\left({(d-1)/ 2}\right)}
\,C_T\,.
\labell{Xder4}
 \eeqa
Of course, the integral in the first line contains a divergence at $r=0$ where the insertion points of the 
two energy-momentum tensors collide. However, recall that the final result must be independent
of the precise choice of the insertion point $x_0^\al$. Hence to regulate the singularity, we could instead evaluate eq.~\reef{Xder} with
$x_0^\al$ shifted slightly away from $\taue=0$.  A simpler approach is to simply evaluate the integral above using dimensional regularization, which yields
 \be
 \int_0^R \frac{dr}{r^{d+2}}\,\frac{R^2-r^2}{2R^2} 
=\frac{1}{d^2-1}\,\frac{1}{R^{d+1}}~.
 \labell{delh}
 \ee
We have explicitly verified that both approaches lead to the same result for eq.~\reef{Xder4}. Finally it
is straightforward to show that the coefficient appearing in eq.~\reef{Xder4} precisely matches the coefficient
appearing in eq.~\reef{interest1}. Hence we have established that the latter is a universal result that applies for twist
operators in any CFT.

The previous analysis is easily extended to higher derivatives of the conformal weight. In particular, eq.~\reef{Xder}
generalizes to
 \bea
 h_{n,1}\equiv\del_n h_n|_{n=1}&=&-{2\pi R^d \over (d-1)T_0}  \langle H\ T_{\taue \taue}(x_0) \rangle_c ~,
 \non
h_{n,2}\equiv \del_n^2 h_n|_{n=1}&=&{2\pi R^d \over (d-1)T_0^2} \Big(\langle H \, H
\ T_{\taue \taue}(x_0)\rangle_c-2\,T_0\,  \langle H\ T_{\taue \taue}(x_0)\rangle_c\Big)~,
 \labell{scalder}\\
 h_{n,k}\equiv\del_n^k h_n|_{n=1}&=&(-1)^{k}{2\pi R^d \over (d-1)T_0^k} \Big(\langle  \underset{k}{\underbrace{H \cdots H} }\ T_{\taue \taue}(x_0)\rangle_c
 -k\,T_0\,  \langle \underset{k-1}{\underbrace{H\cdots H } }\ T_{\taue \taue}(x_0)\rangle_c\Big)~, ~ k\geq2~,
 \nonumber
 \eea
where as before, the expectation values on the right-hand side are the connected parts of the thermal expectation values on the hyperbolic space evaluated at the temperature $T_0$. With these expressions, we can construct a Taylor series for the conformal dimension
around $n=1$, \ie 
\beq
h_n=\sum_{k=1}^\infty \frac{1}{k!}\,h_{n,k}\,(n-1)^k
\labell{series}
\eeq
Note that the series above begins with $k=1$ because, as is evident from eq.~\reef{heavy}, $\lim_{n\to1}h_n=0$.
Therefore eq.~\reef{scalder} shows that this expansion is determined by the correlation functions of the energy-momentum tensor. Our previous analysis provided a universal expression fixing $h_{n,1}$ in terms the central charge $C_T$. For general
$k$, $h_{n,k}$ will be determined by the $(k+1)$- and $k$-point correlators of the stress tensor
and so these expressions will depend on the 
details of the underlying CFT, \eg on the full spectrum of primary operators.\footnote{Of course, $d=2$ is an exception to this general rule. In this case, the relevant expressions are all completely determined by the central charge $c$. In fact, given
the full expression in eq.~\reef{dim}, one finds $h_{n,k}=\frac{c}{12}\Big((-)^{k+1}\Gamma(k+1)+\delta_{k,1}\Big)$.
It is straightforward to confirm that this result agrees with eq.~\reef{interest1} for $k=1$ after identifying $c=2\pi^2\,C_T$. Further for $k=2$, agreement is found with eq.~\reef{q2h}
after substituting in $d=2$ and using eq.~\reef{central}.}
However, to evaluate the second derivative above, we only need the three- and two-point functions,
which are both completely fixed by conformal invariance.  Hence $h_{n,2}$ also has a universal form which depends
on the (three) parameters appearing in the three-point correlator of the stress tensor \cite{Osborn0,Erdmenger0}.
Therefore we turn to deriving this universal expression next.

Following the previous discussion, the correlator implicitly appearing in $h_{n,2}$ is 
\be
\langle T_{\taue \taue}(\taue{}_1,\vec{u}_1)\, T_{\taue \taue}(\taue{}_2,\vec{u}_2)\, T_{\taue \taue}(\taue{}_3,\vec{u}_3)  \rangle_c\,,
\labell{cthree}
\ee
where $\vec{u}_i$ denotes both the radius and the angles at which each of these insertions is positioned on $H^{d-1}$.
However, as before, the results will be independent of the precise choice made for the time slices for the first two stress tensors, which appear
in the Hamiltonians in eq.~\reef{scalder}, and for the position of the third stress tensor. Hence, it will be convenient to
set $\taue{}_1=0=\taue{}_2$, with which these two insertions will be mapped to the slice $\te=0$ and $r=R\tanh u/2\le R$ in flat space, \ie within the entangling surface. However, we will choose $\taue{}_3=\pi\,R$, which also maps the third insertion to $\te=0$ but with $r=R\coth u/2\ge R$, \ie outside of the entangling surface. Further, we will take $u_3\ll1$ below which will
correspond to a limit where $r_3\gg R$. Now in analogy to eq.~\reef{2ptensor}, we map eq.~\reef{cthree} to the corresponding flat space correlator with
\beqa
 &&\langle T_{\taue \taue}(0,\vec{u}_1) \, T_{\taue \taue}(0,\vec{u}_2) \, T_{\taue \taue}(\pi R,\vec{u}_3)  \rangle_c
=\prod_{i=1}^{2}\[ \frac1{\Omega^{d-2}}\,\(\frac{\del\te}{\del\taue}\)^2\]_{\tau_i=0,\vec{u}_i} 
\ \[ \frac1{\Omega^{d-2}}\,\(\frac{\del\te}{\del\taue}\)^2\]_{\tau_3=\pi R,\vec{u}_3} 
\nonumber\\
 &&\qquad
 \times\ \langle T_{\te\te}(\te=0,\vec{r}_1) \, T_{\te\te}(\te=0,\vec{r}_2) \, T_{\te\te}(\te=0,\vec{r}_3) \rangle\Big|_{r_{1,2}=R\tanh u_{1,2}/2;\,r_{3}=R\coth u_{3}/2} ~.
\labell{3ptensor}
\eeqa
Now the first two factors can be simplified using eq.~\reef{very} and similarly, eq.~\reef{CReq} yields
 \be
\frac{\del\te}{\del\taue}\Big|_{\taue=\pi R}=\frac{R^2-r^2}{2\,R^2}=-\frac1\Omega\Big|_{\te=0,r\ge R}~.
\labell{very2}
 \ee
Combining these results, the three-point contribution to $h_{n,2}$ becomes
\beqa
\langle H \, H
\ T_{\taue \taue}(\taue{}_3=\pi R,u_3)\rangle_c &=&
\(\frac{r_3^2-R^2}{2\,R^2}\)^d\ \prod_{i=1}^2 \[\int d\Omega_i\,\int_0^R \!dr_i\,r_i^{d-2}\,
\frac{R^2-r_i^2}{2\,R^2} \]
\labell{forty33}\\
&&\qquad\quad\times\ \langle T_{\te\te}(\te=0,\vec{r}_1) \, T_{\te\te}(\te=0,\vec{r}_2) \, T_{\te\te}(\te=0,\vec{r}_3) \rangle
\nonumber
\eeqa
where the three-point function on the right-hand side is evaluated in $R^d$ (with $r_3>R$). As in eq.~\reef{Xder3},
we also observe that the standard Hamiltonians on the left-hand side  have become entanglement Hamiltonians
for the spherical region on the right-hand side. 

Now we may employ the results of  \cite{Osborn0,Erdmenger0} which give the three-point function of the stress tensor in $R^d$. Recall that this correlator has a universal form that is completely fixed by conformal invariance, tracelessness of the stress tensor and energy conservation, up to three constant parameters which characterize the underlying CFT. The resulting expression is 
quite complicated in general and so we simplify our calculation by using the remaining freedom in choosing $\vec{r}_3$, the position of the third stress tensor. In particular, if we choose $r_3\gg R\ge r_{1,2}$ (or on the hyperbolic space, $u_3\ll1$), the three-point correlator becomes
 \be
 \langle T_{\te\te}(0,\vec{r}_1) \, T_{\te\te}(0,\vec{r}_2) \, T_{\te\te}(0,\vec{r}_3) \rangle \simeq {K\over |\vec r_1- \vec r_2|^d \, r_3^{2d}}~,
 \ee
where
 \be
 K={8(d-1)(d-2)\aha-2d\,\bha -(5d-4)\cha\over d^2}~,
 \ee
with $\aha$, $\bha$ and $\cha$ being the parameters characterizing the CFT.\footnote{Here we are adopting the parametrization of the three-point function of the stress tensor in \cite{Osborn0}, in terms of $\aha,\bha,\cha$. We note that a slightly different parametrization is introduced in \cite{Erdmenger0}, which is also widely used, \eg \cite{gb1,qtop2}. The parameters there are often denoted $\A,\B,\C$ and the relation between these two sets of parameters is given by:
$\A = 8 \aha$, $\B = 8 (2 \aha + \bha)$, $\C = 2 \cha$.} Substituting this expression into eq.~\reef{3ptensor} yields
\beqa
\langle H \, H
\ T_{\taue \taue}(\taue{}_3=\pi R,u_3= 0)\rangle_c &=&
\(\frac{1}{2\,R^2}\)^d\ \prod_{i=1}^2 \[\int d\Omega_i\,\int_0^R\!dr_i\,r_i^{d-2}\,
\frac{R^2-r_i^2}{2\,R^2} \]\ {K\over |\vec r_1- \vec r_2|^d }
\nonumber\\
&&={K\over (2R)^{d+2}}\ I
 \labell{forty44}
\eeqa
where we have written the remaining integral as
 \be
 I=\prod_{i=1}^2 \[\int d\Omega_i\,\int_0^1 \! dx_i \, x_i^{d-2} (1-x_i^2)\]
 {1\over  |\vec x_1- \vec x_2|^d } ~.\labell{integral0}
 \ee
with $x_i=r_i/R$. We evaluate this integral in Appendix \ref{integralx} and with this result, eq.~\reef{forty44} becomes 
 \be
  \langle  H\, H \, T_{\taue \taue}(\taue{}_3=\pi R,u_3= 0)\rangle_c=
-\frac{2\,\pi^{d-2}}{d(d+2)\,\Gamma(d-1)}\,{ K\over R^{d+2}}~.
 \ee
Combining this correlator with eqs.~\reef{Xder4} and \reef{scalder}, we finally obtain
 \beqa
 \del_n^2 h_n|_{n=1}&=&- \frac{16\,\pi^{d+1}}{d^2(d+2)\, \Gamma(d+1)} \, \big(8(d-1)(d-2)\aha-2d\,\bha -(5d-4)\cha\big)
 +2\,h_{n,1}
 \labell{q2h} \\
&=& -\frac{16\pi^{d+1}}{d^2\,\Gamma(d+3)} \left(2  (d-2) (3d^2-3d-4) \aha - 2 d (d-1)\bha -  (3d-4) (d+1) \cha \right)
\nonumber
 \eeqa
where we have simplified the final expression using \cite{Osborn0,Erdmenger0}
\beq
C_T= \frac{4\Omega_{d-1}}{d (d+2)}\,\( (d+3)(d-2) \aha-2 \bha - (d+1)\cha \)\,.
 \labell{central}
 \eeq

\subsection{Comparison with \cite{Perlmutter0}} \labell{eric}

Recently, \cite{Perlmutter0} considered an expansion of the \ren entropy $S_n$ in the vicinity of $n=1$ and found that the first derivative had a universal form similar to eq.~\reef{interest1}. Hence it is instructive to compare our results with the expansion in \cite{Perlmutter0}. As argued in \cite{renyi}, Renyi entropy for 
a spherical region in flat space can be written as
\be
S_n={n\over 1-n}{1\over T_0} [ F(T_0) - F(T_0/n) ]={n\over 1-n} \beta_0 F(\beta_0)-{1\over 1- n} \beta_n F(\beta_n)~,
\labell{older0}
\ee
where $F(\beta)$ is the free energy of the CFT at temperature $T=1/\beta$ on the hyperbolic background $R\times H^{d-1}$. Further $\beta_0\equiv 1/T_0=2\pi R$ and $\beta_n\equiv n/T_0=2\pi R\,n$ --- compare to eq.~\reef{ren3}. In particular, expanding this result in the vicinity of $n=1$ or equivalently around $\beta_0$, yields
\beqa
S_n&=&{n\over 1-n}\beta_0 F(\beta_0)-{1\over 1-n}\(\beta_0 F(\beta_0)+(\beta_n-\beta_0)\,\left.\frac{\del\ }{  \del \beta_n}\!\(\beta_n F(\beta_n)\)\right|_{\beta_n=\beta_0} \right.
\labell{hotwax} \\
&&\qquad\qquad\left.
+{1\over 2} (\beta_n-\beta_0)^2 \left.\frac{\del^2\ }{  \del \beta_n^2}\!\(\beta_n F(\beta_n)\)\right|_{\beta_n=\beta_0}
+{1\over 6} (\beta_n-\beta_0)^3 \left.\frac{\del^3\ }{  \del \beta_n^3}\!\(\beta_n F(\beta_n)\)\right|_{\beta_n=\beta_0} +\cdots\)
 \nonumber
\eeqa
Now this expression may be simplified using $\beta_n-\beta_0=(n-1)\,\beta_0$ and
\be
\left. {\del \beta_nF(\beta_n)  \over \del\beta_n}\right|_{\beta_n=\beta_0}=R^{d-1} V_{\Sigma}\ \E(\beta_0)~,
\labell{simpl99}
\ee
where $R^{d-1} V_{\Sigma}$ denotes the (regulated) volume of the hyperbolic space\footnote{Here, we are using the notation
introduced in \cite{renyi}.} $H^{d-1}$ and, as before, $\E$ denotes the energy density.
Thus eq.~\reef{hotwax} becomes
\be
 S_n=S(\beta_0) +  2\pi R^d\,V_\Sigma \[{1\over 2} (n-1) \beta_0 {\del \E(\beta_0)\over \del \beta_0}+
{1\over 6} (n-1)^2 \beta_0^2 {\del^2 \E(\beta_0)\over \del \beta_0^2}+\cdots\]~,
\labell{hotwax2}
\ee
where $S(\beta_0)$ is the thermodynamic entropy on $R\times H^{d-1}$ at temperature $T=T_0$, which equals the entanglement entropy across the sphere of radius $R$ in flat space \cite{circle4}. Now if we examine eq.~\reef{heavy}, we see that $h_n$ can be written in terms of a similar expansion with derivatives of the energy density
 \be
h_n=-\frac{2\pi R^d}{d-1}\,\( (n-1)\,\beta_0{\del \E(\beta_0)\over \del \beta_0} +\frac{(n-1)^2}{2}\(\beta_0^2 {\del^2 \E(\beta_0)\over \del \beta_0^2}
+2\beta_0 {\del \E(\beta_0)\over \del \beta_0}\)+\cdots \)\,.
 \labell{heavy2}
 \ee
Hence comparing the leading coefficient in the two expansions, we find
\beq
 \left.\del_n S_n \right|_{n=1}=-{d-1\over 2}\,V_{\Sigma}\, \left.\del_n h_n \right|_{n=1}~.
\labell{hotwax1}
\eeq
Similarly, comparing the expansions at higher orders yields
\beqa
\left.\del^2_n S_n \right|_{n=1}&=&-{d-1\over 3}\,V_{\Sigma}\, \(h_{n,2}-2\,h_{n,1}\)~,
\labell{hotwax3}\\
\left.\del_n^k S_n \right|_{n=1}&=&-{d-1\over k+1}\,V_{\Sigma}\, \(h_{n,k}-k\,h_{n,k-1}\)~,
\nonumber
\eeqa
where we are using the notation introduced in eq.~\reef{scalder} here, \ie $h_{n,k}=\left.\del^k_n h_n \right|_{n=1}$.
In particular, substituting eq.~\reef{interest1} into eq.~\reef{hotwax1}, we recover the result  derived in \cite{Perlmutter0} 
\be
  \del_n S_n\big|_{n=1}=-  \pi^{\frac{d}{2}+1}\,
 \frac{ (d-1)\,\Gamma \( {d}/{2}\)}{\Gamma(d+2)}\,V_\Sigma\ C_T\,.
 \labell{interest22}
\ee
Further, using eqs.~\reef{q2h} and \reef{central}, we could express the second derivative $\del_n^2 S_n\big|_{n=1}$ in terms of the
CFT parameters $\aha$, $\bha$ and $\cha$. However, as with the expansion of $h_n$, the higher derivatives of the \ren entropy would not have a simple universal
form.

\section{Explicit examples} \labell{compare}

In this section, we explicitly evaluate the conformal weight $h_n$ in several theories using eq.~\reef{heavy}. Given the expression for $h_n$, we can
calculate the derivatives $h_{n,1}$ and $h_{n,2}$ and then compare the results to eqs.~\reef{interest1} and \reef{q2h}. In this way, it is first observed for a variety of
holographic models \cite{renyi} that the first derivative $h_{n,1}$ had a simple universal form and the latter then motivated the general proof for a generic CFT, which
we presented in the previous section. Since
this proof extends to second derivatives, this new expression in eq.~\reef{q2h} can also be compared with the results for $h_{n,2}$ obtained from holography in sections \ref{love} and \ref{QT}. Free fields, \eg a massless fermion or a conformally coupled scalar, are another case where the relevant computations can be explicitly performed. We present the results of our comparison for these free theories in section \ref{free1}, while the details of the heat kernel calculations for the free fields appear in appendix \ref{appfree}.

\subsection{Holographic Lovelock gravity} \labell{love}

Let us begin with a holographic framework where the bulk is described by Lovelock gravity with up to six-derivatives in $d+1$ dimensions \cite{lovel}.\footnote{For holographic studies of Lovelock gravity, see for example \cite{gb1,gb2,highc}.} The inclusion of the six-derivative terms here extends the results for $h_{n,1}$ in Gauss-Bonnet gravity
already described in \cite{renyi}. The gravitational action is given by
\bea
{I}_{\text{\tiny{Lovelock}}}&&= \frac {1}{2\lp^{d-1}}\int d^{d+1}x \sqrt{-g}\,
\Big[\frac{d(d-1)}{L^2}+R+\frac{L^2\,\lgb}{(d-2)(d-3)}\cL_4
\nonumber\\
&&\qquad\qquad\qquad+ {\frac{L^4\,\mu}{(d-2)(d-3)(d-4)(d-5)}\cL_6} \Big] \,,
\labell{SLV}
\eea
where the two higher curvature interactions take the form
\begin{equation}
\cL_4=R_{abcd}R^{abcd}-4 R_{ab}R^{ab}+R^2\ ,
 \label{euler4}
\end{equation}
\vskip -.9cm
 \bea
\cL_6 &=& 4\, R_{\mu\nu}^{\,\,\,\,\,\,\rho\sigma}
R_{\rho\sigma}^{\,\,\,\,\,\,\tau\chi} R_{\tau\chi}^{\,\,\,\,\,\,
\,\,\mu\nu}-8\, R_{\mu\,\,\,\nu}^{\,\,\,\rho\,\,\,\,\sigma}
R_{\rho\,\,\,\,\sigma}^{\,\,\,\tau\,\,\,\chi}
R_\tau{}^\mu{}_\chi{}^{\nu} -24\, R_{\mu\nu \rho\sigma} R^{\mu\nu
\rho}{}_{\tau} R^{\sigma\tau} +3\, R_{\mu\nu\rho\sigma}
R^{\mu\nu\rho\sigma} R
 \nonumber\\
&&\qquad\quad
 +24\,R_{\mu\nu\rho\sigma} R^{\mu\rho}R^{\nu\sigma}+16\, R_\mu^{\,\,\nu}
R_\nu^{\,\,\rho} R_\rho^{\,\,\mu} -12\, R_\mu^{\,\,\nu} R_\nu^{\,\,\mu}
R + R^3\,. \labell{euler6}
 \eea
The scale $L$ appearing in the action \reef{SLV} is related to the radius of curvature $\tilde{L}$ of the
corresponding AdS vacuum by
\be
\tilde{L}^2= L^2/\fin,\qquad {\rm where}\ \ 1-\fin +\lgb \fin^2 + \mu \fin^3 =0\,.
\ee
Now following \cite{renyi}, the thermodynamic properties of the dual CFT in the background $R\times H^{d-1}$ are determined by studying hyperbolic AdS black holes.
In particular, the temperature of such an AdS black hole can be written as

\be
T= \frac{1}{2\pi R\, x} \left (1 + \frac{d}{2 \fin} \frac{x^6 - 
     \fin x^4 + \lgb \fin^2 x^2 + \mu \fin^3}{x^4 - 2 \lgb \fin  x^2 - 3 \mu \fin^2 } \right)
\label{tempp}
\ee
where $x=r_\mt{H}/\tilde{L}$ with $r_\mt{H}$ being the horizon radius. However, recall that we are particularly interested in temperatures
\be
T= T_0/n=\frac{1}{2\pi R \,n}\,,
\labell{ttnn}
\ee
and with this choice, eq.~\reef{tempp} determines $x=x_n$ for a given $n$.

Now the energy density can be determined from the black hole solutions and then the scaling dimension $h_n$ of the twist operators can be obtained using eq.~\reef{heavy}.
However, the calculations are simpler if the latter is re-expressed the latter in terms of the entropy \cite{renyi}
\be
h_n=\frac{2\pi R\,n}{(d-1)\,V_{\Sigma}}\,\int_{x_n}^1 dx\  T(x)\,\frac{dS(x)}{dx} \,.
\labell{heavy4}
\ee
The horizon entropy can be evaluated with Wald's formula \cite{waldent} and
for completeness, we give the final expression for the hyperbolic AdS black holes in Lovelock gravity
\be
S(x) = 2 \pi   \left(\frac{\tilde{L}}{\lp}\right)^{d-1}\!V_\Sigma\   x^{d - 1} \left(1 - 2 \frac{d - 1}{d - 3}\, \frac{\lgb \fin} {x^2} -
    3 \frac{d - 1}{d - 5}\, \frac{\mu \fin^2}{x^4}\right)\,.
\label{entropyy}
\ee
Eq.~\reef{heavy4} then yields
\be
h_n =  \pi\,\left(\frac{\tilde{L}}{\lp}\right)^{d-1}\! n\, x_n^{d-6} (x_n^2-1) \left(\mu \fin^2 + x_n^4 (\lgb\fin
+ \mu \fin^2-1) + x_n^2 (\lgb \fin + \mu \fin^2)\right)\,.
\ee
Since we will
be taking the limit $n\to1$ at the end, we need only to obtain a perturbative
solution for $x_n$ around $n=1$. This can be readily solved, giving
\bea\label{solx}
x_n&=& 1+ \alpha_1 (n-1) + \alpha_2 (n-1)^2 +\cdots \qquad{\rm with}\labell{hotwax88}\\
\alpha_1&=& -\frac{1}{ d-1}\,,\qquad \alpha_2 = -\frac{d (3 - 10 \lgb\fin - 21 \mu \fin^2 -2
   d (1-2 \lgb\fin -3  \mu \fin^2))}{2 (d-1)^3 (
   1-2 \lgb\fin - 3 \mu \fin^2 )}\,, \nonumber 
\eea
which in turn, yields 
\beqa
h_{n,1}&=&\frac{2\pi}{d-1}\,\left(\frac{\tilde{L}}{\lp}\right)^{d-1}  \left(1-2\lgb\fin-3\mu \fin^2 \right)\,,
\labell{gmmah}\\
h_{n,2}&=& -\frac{2\pi}{(d-1)^3}\,\left(\frac{\tilde{L}}{\lp}\right)^{d-1}
 \bigg(1-6\lgb \fin - 15 \mu \fin^2 - 4d(1- 4\lgb \fin - 9\mu \fin^2) \nonumber \\
&&\qquad\qquad \qquad\qquad \qquad\qquad +2d^2 (1-2\lgb \fin - 3\mu\fin^2) \bigg)\,.
\nonumber
\eeqa

Finally we need rewrite the above expressions in terms of parameters from the boundary CFT. Three such CFT parameters
which are readily determined in terms of the free parameters in the gravitational theory, \ie $\lgb$, $\mu$ and $\tilde{L}/\lp$, 
are \cite{gb1}
\bea 
C_T &=& \frac{(d +1)\Gamma[d + 1]}{(d - 1) \pi^{d/2} \Gamma[d/2]} \left(\frac{\tilde{L}}{\lp}\right)^{d-1}  (1 - 2 \lambda \fin - 3 \mu \fin^2)\,,
\labell{relations} \\
t_2 &=& \frac{4d(d -1)(\lambda \fin + 3 \mu \fin^2)}{ (d - 2) (d - 3)(1 - 2 \lambda \fin - 3 \mu \fin^2)}\,,\qquad t_4=0\,.
\eea
where $t_2$ and $t_4$ are constants appearing in certain thought experiments involving the measurement of energy fluxes \cite{diego} --- see also \cite{gb1,qtop2}. 
Now for any CFT, these coefficients $C_T$, $t_2$ and $t_4$ are related to the parameters $\aha$, $\bha$, $\cha$, described in the previous section, with
\cite{diego,gb1}
\bea
C_T&=& 4\Omega_{d-1}\,\frac{(d-2)(d+3)\,\aha-2\, \bha - (d+1)\,\cha }{d (d+2)}\nonumber \\
t_2&=& \frac{2 (d+1)  (
   (d-1)(d^2+8d+4)\,\aha+3 d^2\,\bha-d (2 d+1)\,\cha)}{d ((d-2)(d+3)\,\aha-2\, \bha - (d+1)\,\cha)} \labell{rosetta} \\
t_4&=&-\frac{( d+1) ( d+2) (3(2d+1)(d-1)\,\aha+2d^2\,\bha -d (d+1)\,\cha)}{
 d ((d-2)(d+3)\,\aha-2\, \bha - (d+1)\,\cha)}\,.\nonumber
\eea
Since $t_4$ vanishes in Lovelock gravity, there is a constraint which allows us to eliminate one of the parameters
$\aha$, $\bha$ and $\cha$ in favour of the other two.  In particular, we write
\be
\cha = \frac{3(2d+1)(d-1)\,\aha+2 d^2 \,\bha }{d (d + 1)}\,.  \labell{cons}
\ee
Taking eqs.~\reef{rosetta} and \reef{cons} into account, eq.~\reef{gmmah} becomes
\beqa
h_{n,1}&=& 
\frac{16 \pi^{
d+1}}{d \Gamma(d+3)}\,\((d^2-6d + 3)\aha-2 d \bha  \)  \,,
\labell{gmmai}\\
h_{n,2}&=& -\frac{16 \pi^{d+1} }{d^3 \Gamma(d+3)}\,\left((6d^4-36d^3+37d^2+ 13d-12) \aha-2d^2  (4d-5)  \bha 
     \right)\,,
\eeqa
which is in perfect agreement with our CFT results, \ie with eqs.~\reef{interest1} and \reef{q2h} when we substitute in eq.~\reef{cons}.

\subsection{Holographic quasi-topological gravity} \labell{QT}

To explore more general holographic CFT's, \ie with $t_4\ne0$, we also consider quasi-topological gravity \cite{qtop2,qtop1} with four boundary dimensions,
as in \cite{renyi}. For completeness, the action is given by
\be
{I}_{\text{\tiny{quasi-top}}}= \frac {1}{2\lp^{3}}\int d^5x \sqrt{-g}
\Big[\frac{12}{L^2}+R+\frac{1}{2}\,L^2\,\lgb\,\cL_4+ \frac{7}{8} L^4\,\mqt\,\mathcal{Z}_5' \Big]
\labell{quasitop}
\ee
where 
\bea
\mathcal{Z}_5' = && R_{\mu\nu}{}^{\rho\sigma}R_{\rho\sigma}{}^{\alpha\beta}R_{\alpha\beta}{}^{\mu\nu} + \frac{1}{14} (21 R_{\mu\nu\rho\sigma}R^{\mu\nu\rho\sigma}R - 120R_{\mu\nu\rho\sigma}
R^{\mu\nu\rho}{}_{\alpha}R^{\sigma\alpha} \nonumber \\
&&+ 144 R_{\mu\nu\rho\sigma} R^{\mu\rho} R^{\nu\sigma} + 128 R_{\mu}{}^{\nu}R_{\nu}{}^{\rho}R_{\rho}{}^{\mu} - 108 R_{\mu}{}^{\nu}R_{\nu}{}^{\mu} R + 11 R^3)\,.
\eea

The expressions for the
temperature and entropy of a hyperbolic AdS black hole are precisely as in eqs.~(\ref{tempp}) and (\ref{entropyy}), respectively, upon making the replacement $\mu \to \mqt$ and taking $d=4$.
Since the formula for the temperature stays the same, the perturbative solution for $x_n$ in eq.~(\ref{solx}) also remains unchanged. 
Similarly, the expression for $C_T$ in eq.~(\ref{relations}) is the same in the present case of quasi-topological gravity.
The new ingredients are the values for the parameters $t_2$ and $t_4$, for which we have \cite{qtop2}
\bea
t_2 &=& 24 \, \frac{\lgb \fin - 87 \mqt \fin^2}{1 - 2 \lgb \fin - 3 \mqt\fin^2}\,, \nonumber \\
t_4 &=& 3780\, \frac{\mqt \fin^2}{1 - 2 \lgb \fin - 3 \mqt \fin^2}\,.
\eea
Again collecting these results and applying eq.~\reef{rosetta}, we arrive at
\beqa
h_{n,1}&=&\frac{\pi^5}{180}\,(14\aha-2\bha-5\cha)\,,
\labell{gmmaj}\\
h_{n,2}&=&-\frac{\pi^5}{90}\,(16\aha-3\bha-5\cha)\,.
\eeqa
which are again in perfect agreement with eqs.~\reef{interest1} and \reef{q2h} with $d=4$.

\subsection{Free fields} \labell{free1}

In this subsection, we consider making a comparison to  eqs.~\reef{interest1} and \reef{q2h} in the special cases of free massless fermions and free conformally coupled scalar fields. As expected, a match is found for $h_{n,1}$ and $h_{n,2}$ calculated with heat kernel techniques for these free fields. We present here only our final findings, whereas the details of the computations are relegated to Appendix \ref{appfree}.

As shown in \cite{Osborn0}, the free fields under consideration satisfy
\be
C_T={d\over \Omega_{d-1}^2}\, \left({n_s\over d-1}  + {n_f\over 2}\right)~,
\labell{CT}
\ee
where $n_s$ and $n_f$ denote the number of (massless) degrees of freedom contributed by the scalar and fermion fields, respectively. In particular for a real scalar field, $n_s=1$, whereas for a Dirac fermion,\footnote{Here, ${\lfloor{d/ 2}\rfloor}$ denotes the integer part of $d/2$.} $n_f=2^{\lfloor{d\over 2}\rfloor}$. Similarly, the parameters appearing in the three-point correlator of the stress tensor take the form \cite{Osborn0}
\bea
&&\aha=\frac{d^3}{8\Omega_{d-1}^3(d-1)^3}\,n_s\, , \quad
\bha=-\frac{d^2}{8\Omega_{d-1}^3}\left(\frac{d^2}{(d-1)^3}\,n_s+{n_f\over 2}  \right)\, , 
\nonumber\\
&&\qquad\qquad\quad
\cha=-\frac{d^2}{8\Omega_{d-1}^3}\left(\frac{(d-2)^2}{(d-1)^3}\,n_s+ n_f \right)\, .
\labell{fchrg}
\eea
Hence, with the above expressions, eq.\reef{interest1} takes the following simple form
 \be
 h_{n,1}
=\frac{d\,\Gamma(d/2)^3}{2\pi^{(d-2)/2}\Gamma(d+2)}\, \left({n_s\over d-1} + {n_f\over 2}\right)~, \labell{hothouse}
 \ee
while eq.~\reef{q2h} becomes
 \be
 h_{n,2}
=-\frac{d\,\Gamma(d/2)^3}{4\pi^{(d-2)/2}\Gamma(d+3)}\, \left(
\frac{11d^4-33d^3+16d^2+28d-16}{(d-1)^3}\,n_s
 + 2(2d^2-d-2)\,n_f\right)~. \labell{hothouse2}
 \ee

On the other hand, in the case of free fields, $h_n$ can be evaluated in full generality by means of heat kernel methods, as we describe in Appendix \ref{appfree}. By explicit calculations from this approach then, we have verified that  first derivative of the scaling dimension so obtained agrees with eq.~\reef{hothouse} in various spacetime dimensions up to $d=14$. We were also able to explicitly verify that this approach reproduces eq.~\reef{hothouse2} for massless fermions in dimensions up to $d=12$. Unfortunately, we must report that for the
conformally coupled scalar, there was a discrepancy between $h_{n,2}$ as evaluated using the heat kernel results and as given in eq.~\reef{hothouse2}. We return to this issue
in section \ref{discuss}.

\section{OPE of spherical twist operators} \labell{OPE}

In section \ref{twist}, we described how the conformal mapping \reef{change} between
$S^1\times H^{d-1}$ and the $n$-fold cover of $R^d$ could be applied to evaluate the
expectation value of the stress tensor in the presence of a spherical twist operator as in eq.~\reef{stressfin}, 
\ie
 \be
\langle T_{\al\beta}\ \sigma_n \rangle =\Omega^{d-2}\,\frac{\partial X^\mu}{\partial
x^\al}\,\frac{\partial X^\nu}{\partial x^\beta}\,\Big(\langle T_{\mu\nu}(T_0/n)\rangle
-\langle T_{\mu\nu}(T_0)\rangle\Big)\, .
 \labell{stressfin2}
 \ee
where the conformal factor $\Omega$ is given by eq.~\reef{factx}. We examined this result
in the limit where the stress tensor was brought very close to the twist operator in order to determine the
conformal dimension of twist operators in the CFT in terms of the thermal energy density in the
hyperbolic background, as given in eq.~\reef{heavy}. In this section, we consider the opposite
limit where the stress tensor is taken very far from the twist operator, \ie  $T_{\al\beta}$ is inserted at
a large radius $r$ with $r\gg R$.

In investigating any twist operator enclosing some finite region with long wavelength probes, such
as in the correlator described above, the twist operator $\sigma_n$ can be approximated by
a sum of local operators $\mathcal{O}_p$ and their descendants (indexed by $k$ below) \cite{shifman}\footnote{Since
$\sigma_n$ is a nonlocal operator, there are certain ambiguities here. That is, the precise
form of the expansion coefficients $c^n_{\,p,k}$ will depend on the choice of the scale $R$ and of the reference point at which the
operators $\mathcal{O}_p^k$ are inserted. In eq.~\reef{sigmaope}, we implicitly choose $R$ is the radius of the sphere and the reference point
as the center of the spherical region enclosed by the twist operator.} 
\begin{equation}
\sigma_n = \langle \sigma_n\rangle \left( 1+\sum_{p,k} R^{\Delta_{p,k}}\ c^n_{\,p,k}\ \mathcal{O}_p^k\right)\,,
\labell{summ88}
\end{equation}
where $R$ is some (macroscopic) scale characterizing the size of $\sigma_n$ and the $\Delta_{p,k}$ are the conformal dimensions
of the operators $\mathcal{O}_p^k$.
Two comments on this expansion are: The operators $\mathcal{O}_p$ may be conformal primaries in a single
copy of the CFT, but in general they will be products of two or more such operators inserted at the same
point but in different copies of the CFT, \ie
on different sheets of the $n$-fold cover. In $d=2$ dimensions, the twist operators are themselves local operators but 
calculating the \ren entropy of an interval requires the insertion of two twist operators, one at each end of the interval.
Hence in this case, \ie $d=2$, the above expansion corresponds to the operator product expansion coming
from the fusion of the two twist operators.\footnote{The fusion rules of a pair twist operators have been computed in
specific models, such as the Ising model, for example in \cite{extra1,fuse1}.} Hence following the common nomenclature, \eg \cite{shifman},
we refer to eq.~\reef{summ88} as the `operator product expansion' (OPE) of a single twist 
operator on a closed surface in higher dimensions.

Now we can use the expectation value \reef{stressfin2} of the stress tensor in the presence of a spherical twist operator in a CFT
to learn something about the corresponding OPE \reef{summ88}.
If the OPE is used to replace $\sigma_n$ in this expectation value, the only nonvanishing contribution will come from 
the descendants of the identity, \ie conformal invariance dictates that $\langle T_{\al\beta}\,\mathcal{O}_p^k\rangle=0$ for other operators. 
Hence for large separations, the leading contribution is\footnote{Recall that the correlator $\langle T_{\al\beta}\ \sigma_n \rangle$ is implicitly
normalized by dividing by $\langle \sigma_n \rangle$ but we have been leaving this normalization implicit to avoid further clutter. In any event,
this normalization removes the factor of $\langle \sigma_n\rangle$ appearing in eq.~\reef{summ88} from the right-hand side of eq.~\reef{sigmaope}.}
\begin{equation}
\langle T_{\mu\nu}(x)\, \sigma_n \rangle =   R^d\,\varepsilon_n^{\alpha\beta} \  \langle T_{\mu\nu}(x)\, T_{\alpha\beta}(0) \rangle + \cdots 
\labell{sigmaope}
\end{equation}
where the OPE coefficient takes the form $\varepsilon_n^{\alpha\beta}$, some (traceless and symmetric) polarization tensor that in general depends on the geometry of the
surface operator. The ellipsis above denotes the higher descendants with more insertions of the stress tensor. Hence the leading long-distance behaviour in 
$\langle T_{\mu\nu}(x)\, \sigma_n \rangle$ is controlled by the two-point correlator of the stress tensor, given in eq.~\reef{emt2p}. 

To determine the precise form of $\varepsilon_n^{\alpha\beta}$ for spherical twist operators, we first examine the long-distance behaviour of the expectation value in eq.~\reef{stressfin2}. It will be sufficient to only evaluate
the latter at $t_\mt{E}=0$ and some fixed angles. From eqs. \reef{factx} and \reef{CReq}, it follows then that in the limit $r\gg R$
\be
 \Omega\simeq{2R^2\over r^2}\, , \quad
 \frac{\partial u}{\partial \te}=-{1\over R}\frac{\partial \taue}{\partial r} = 0 \, ,  \quad
 \frac{\partial u}{\partial r}={1\over R}\frac{\partial \taue}{\partial\te}\simeq  -{2 R\over r^2}~.
\ee
Substituting these expressions into eq.~\reef{stressfin2}, we obtain
\beqa
 \langle T_{\te\te}\, \sigma_n \rangle&=&\bigg({2R^2\over r^2}\bigg)^d \Big(\E(T_0)-\E(T_0/n) \Big) ={d-1\over 2\pi n}\bigg({2R\over r^2}\bigg)^d h_n\, ,
  \nonumber\\
  \langle T_{ij}\, \sigma_n \rangle&=&-{g_{ij}\over d-1}\bigg({2R^2\over r^2}\bigg)^d \Big(\E(T_0)-\E(T_0/n) \Big) =-{g_{ij}\over 2\pi n}\bigg({2R\over r^2}\bigg)^d h_n\, ,
\labell{wacko}
\eeqa
where we have used eqs.~\reef{none5}, \reef{none6} and \reef{heavy}, as well as $\sinh u\simeq2R/r$. Here, the indices $i,j$ run over all of the directions orthogonal to $\te$, \ie $r$ and $d-2$ angular coordinates,
and $g_{ij}$ is the flat space metric on these directions, \ie $g_{ij}={\rm diag}(1,r^2,r^2\sin^2\theta,\cdots)$. To simplify the comparison with eq.~\reef{sigmaope}, we adopt Cartesian coordinates on these directions
at this point, in which case we have simply $g_{ij}=\delta_{ij}$. Finally, we note that
in eq.~\reef{wacko}, $T_{\mu\nu}$ is inserted on a single sheet of the $n$-fold cover and therefore we must multiply it by $n$ to in order to produce $\langle T_{\mu\nu}\,\sigma_n\rangle$ with the full stress tensor
of the $n$ copies of the CFT. Hence our final result is
\be
 \langle T_{\te\te}\,\sigma_n\rangle={d-1\over 2\pi }\bigg({2R\over r^2}\bigg)^d\, h_n ~ , \quad\quad \langle T_{ij}\,\sigma_n\rangle=-{\delta_{ij}\over 2\pi }\bigg({2R\over r^2}\bigg)^d\, h_n\, ,
\labell{compar00}
\ee
whereas all other components of $\langle T_{\mu\nu}\,\sigma_n\rangle$ vanish (when we set $\te=0$). Of course, we see here that the long-distance behaviour of the expectation value \reef{stressfin2} can be expressed in terms of the conformal dimension $h_n$ of the twist operators.\footnote{Actually, we might use this long-distance correlator with a spherical
twist operator as an alternate definition of the conformal dimension. Further, let us add that while we have explicitly shown that $h_n$ (as well as $R$) controls the
expectation value $\langle T_{\mu\nu}\,\sigma_n\rangle$ at large and small separations,  in fact, one can easily verify from eq.~\reef{stressfin} that the same is true
of the entire correlator at any separation, \eg see eq.~\reef{almost88}.}

Now we return to eq.~\reef{sigmaope} and substitute eq.~\reef{emt2p} for the two-point correlator of the stress tensor. However, as above, we will restrict our attention to $\te=0$ in which
case the result may be written as
\begin{equation}
\langle T_{\mu\nu}(\te=0,x^i)\, \sigma_n \rangle \simeq   R^d\,\varepsilon_n^{\alpha\beta} \  \frac{C_T}{r^{2d}} \(\delta_{\al\mu}-2\hat r_\al\,\hat r_\mu\)\(\delta_{\beta\nu}-2\hat r_\beta\,\hat r_\nu\)
\labell{sigmaope55}
\end{equation}
where we have used the tracelessness of $\varepsilon_n^{\alpha\beta}$ to simplify the above expression.
We have also defined here the radial unit vector in the $x^i$ directions: $\hat r^\mu\equiv(0,x^i/r)$. Below, it will also
be useful to define the unit vector point along the $\te$ direction: $\hat t^\mu\equiv(1,\vec{0})$.

Let us consider the possible form for $\varepsilon_n^{\alpha\beta}$. The geometry of the spherical entangling surface
demands this polarization tensor must be rotationally symmetric in the $x^i$ directions. Hence most general allowed tensor can be written as
\be
\varepsilon_n^{\alpha\beta}=\al_1\,\delta^{\al\beta}+\al_2\,\hat t^\al\,\hat t^\beta+\al_3\,\hat r^\al\,\hat r^\beta+\al_4\,\(\hat r^\al\,\hat t^\beta+\hat t^\al\,\hat r^\beta\)\,,
\labell{allow33}
\ee
where $\al_{1,2,3,4}$ are all only functions of $r$. Further the tracelessness of the polarization tensor requires that
\be
0=d\,\al_1+\al_2+\al_3\,.
\labell{tracy8}
\ee
Now substituting eq.~\reef{allow33} into eq.~\reef{sigmaope55} yields
\begin{equation}
\langle T_{\mu\nu}(\te=0,x^i)\, \sigma_n \rangle \simeq   \frac{C_T\,R^d}{r^{2d}} \(\al_1\,\delta_{\mu\nu}+\al_2\,\hat t_\mu\,\hat t_\nu+
\al_3\,\hat r_\mu\,\hat r_\nu-\al_4\,\(\hat r_\mu\,\hat t_\nu+\hat t_\mu\,\hat r_\nu\)\)
\labell{sigmaope66}
\end{equation}
Therefore comparing this expression with the results from the conformal mapping in eq.~\reef{compar00}, we find
\be
\al_1= -\frac{2^{d-1}\,h_n}{\pi\,C_T}\,,\quad\al_2=-d\,\al_1\,,\quad \al_3=0=\al_4\,.
\labell{gfire}
\ee
That is, we can write the polarization tensor $\varepsilon_n^{\alpha\beta}$ appearing in the OPE of a spherical twist operator as
\be
\varepsilon_n^{\alpha\beta}=\frac{2^{d-1}\,h_n}{\pi\,C_T}\,\(d\,\hat t^\al\,\hat t^\beta-\delta^{\al\beta}\)\,.
\labell{gfire8}
\ee
Finally we note that the term proportional to $\delta^{\al\beta}$ can be dropped because the trace of the stress tensor
vanishes. Hence the contribution of the stress tensor to the OPE \reef{summ88} reduces to
\be
\sigma_n(0) = \langle \sigma_n\rangle \left( 1+ \gamma\,h_n\,R^{d}\ T_{\te\te}(0) +\cdots \right)\qquad
{\rm with}\ \ \gamma=\frac{2^{d-1}d}{\pi\,C_T}\,,
\labell{summ95}
\ee
for a spherical twist operator of radius $R$ positioned as the origin. We note that the above OPE coefficient depends on the order
of the twist operator only through the appearance of $h_n$, \ie the factor $\gamma$ is independent of $n$.

Of course, a similar analysis using the conformal mapping  in section \ref{twist} could be made to evaluate other terms
in the OPE of a spherical twist operator --- see section \ref{discuss}.

\section{Twist operators near $n=1$} \labell{small}

Twist operators $\sigma_n$ are originally defined for integer $n\ge2$, however, formally we can consider these operators
for arbitrary $n$. Such a continuation was already implicit in section \ref{dimens}, where the conformal dimension $h_n$ was expanded near $n=1$. In this section, we consider a similar expansion for small $(n-1)$ of the twist operators themselves. 
Again, this is a formal expansion which indicates that any correlators involving $\sigma_n$ will behave in a universal manner
in the limit $n\to1$. Note that our discussion here is quite general, \ie it is not limited to spherical entangling surfaces
or to CFT's. Rather our main result in eq.~\reef{toomuch} applies for general entangling geometries and for any quantum field theory.

Our approach will be to begin by considering the correlator of the twist operator $\sigma_n$ with some collection of operators, which we denote collectively  as $\X$.\footnote{In general,
$\X$ may include both local and nonlocal operator insertions, \eg when the underlying QFT is a gauge theory, $\X$ may include Wilson line operators.} 
However, we will restrict our attention
to the case where all of the operators comprising $\X$ are in a single copy of the QFT, say, the first of the $n$ copies. That is, generally the correlators $\langle \sigma_n\,\X\rangle$ are defined 
by inserting the operators $\X$ in the path integral of the QFT on the $n$-fold covering space but we limiting our considerations to the situation
where all of the insertions are made on the first sheet. Next, we construct a new `effective twist operator' $\tsigma_{n}$ which only acts within the first QFT
but reproduces any such correlator, \ie 
\be
\langle \sigma_n\,\X\rangle
\ =\ \langle \tsigma_{n}\,\X\rangle_1
\labell{simple88} 
\ee
where the subscript on the second correlator indicates that this expression is evaluated in the first copy of the QFT. Formally, it is straightforward to define this new operator by integrating out
the other copies of the QFT for which there are no operator insertions in the above correlators. That is, 
\be
\tsigma_n \equiv\langle\,\sigma_n\,\rangle_{\{2,\cdots,n\}}\,,
\labell{simple99}
\ee
where the subscript on the right-hand side indicates that we are performing the path integral over the ($n$--1) copies of the QFT other than the first copy. Now let us consider the Euclidean
path integral representation of the correllator $\langle \sigma_n\,\X\rangle$ with an $n$-fold covering geometry, as illustrated for a simple example in the figure \ref{rilke9}a.  Recall that the $n$-fold cover 
was formulated to give a path integral construction of (integer) powers of the reduced density matrix, as in eq.~\reef{ren2}. Hence when eq.~\reef{simple99} instructs
us to perform the path integral over the $(n-1)$ empty sheets, we can interpret this portion as the path integral representation of the operator $\rhoa^{n-1}$. Therefore the same
correlator can be evaluated within a single copy of the QFT by including an insertion of the latter operator in the region A, as shown in  figure \ref{rilke9}b. Hence we are led to conclude that the
effective twist operator corresponds to\footnote{The same conclusion can be reached with the following formal manipulations:
\be
\langle \sigma_n\,\X\rangle=\Tr\[\rhoa^n\,\X\]=\Tr\[\rhoa\,\rhoa^{n-1}\,\X
\]
=\langle \rhoa^{n-1}\,\X\rangle_1\,,
\labell{dense88}
\ee
where again the subscript on the last correlator indicates that this expression is evaluated in the first copy of the QFT. 
However, for the two intermediate expressions, \ie $\Tr\[\rhoa^n\,\X\]$ and $\Tr\[\rhoa\,\rhoa^{n-1}\,\X\]$, the operator insertions are implicitly limited to be 
within the region A. Of course, the general discussion above had no such restriction.}
\be
\tsigma_n =\rhoa^{n-1}\,.
\labell{simple77}
\ee
Let us re-iterate that it is essential for this argument that all of the operators in $\X$ were from a 
single copy of the QFT or alternatively, that they are all inserted on a single sheet of the $n$-fold covering geometry.
\begin{figure}[h!]
\centering
\subfloat[]{\includegraphics[width=0.6\textwidth]{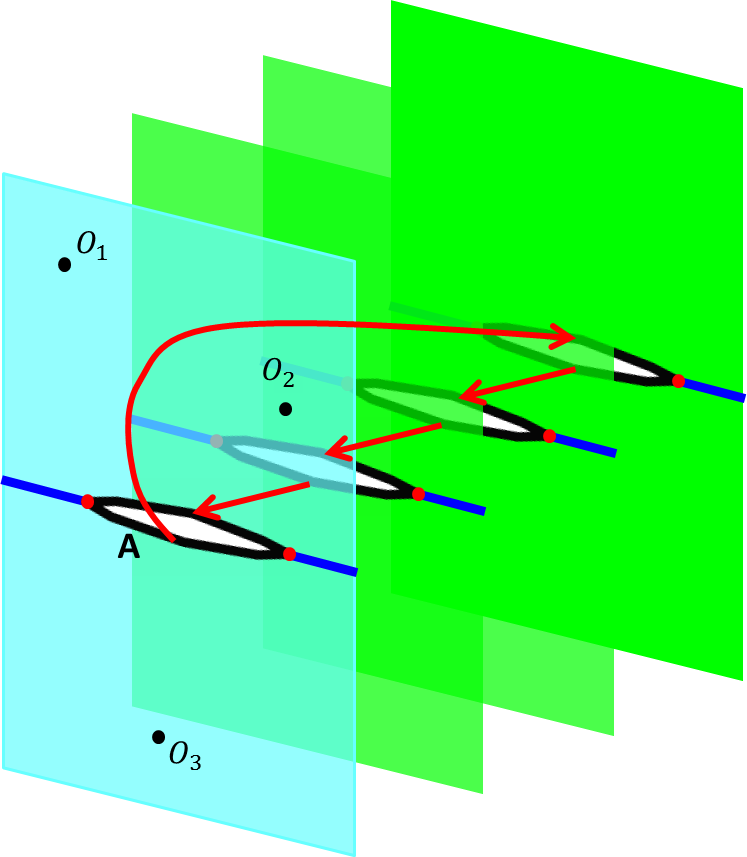}}\qquad\qquad
\subfloat[]{\includegraphics[width=0.288\textwidth]{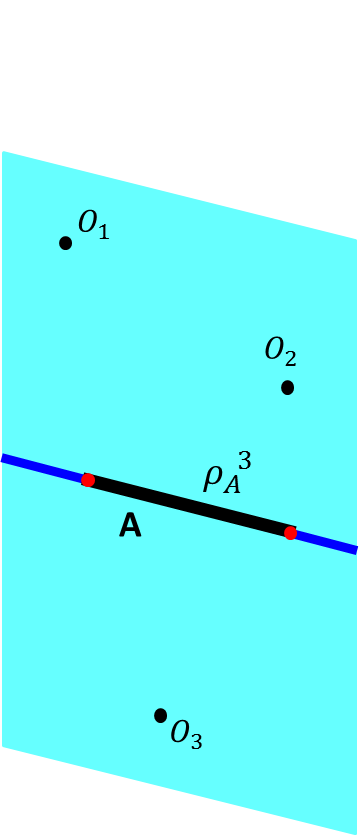}}
\caption{(Colour online) Panel (a) shows the four-fold geometry which would appear in the evaluation of the correlator $\langle\sigma_4\, O_1\,O_2\,O_3\rangle$, where the
operator insertions all lie on the first sheet. The path integral on the next three sheets (shaded green) provides a representation of $\rhoa^{\,3}$. Hence in (b), the
same correlator is evaluated as a correlator on a single sheet with an operator insertion $\rhoa^{\,3}$ in the region A.}\labell{rilke9}
\end{figure}

Finally let us recall that the reduced density matrix appearing in the calculation of the \ren entropy \reef{ren2} can be expressed as
 \be
\rhoa=e^{- \Hm}
\labell{important}
 \ee
for some Hermitian operator $\Hm$. The latter is known as the modular Hamiltonian
in the literature on axiomatic quantum field theory, \eg
\cite{haag}, while it is referred to as the entanglement Hamiltonian in the
condensed matter theory literature, \eg \cite{cmt}. However, we emphasize that generically the entanglement
Hamiltonian is not a local operator and the evolution generated by $\Hm$ would
not correspond to a local (geometric) flow. However, a CFT reduced to a spherical region
provides an exception to this general rule. In this case, we can write $\Hm$ as \cite{circle4}
\be
\Hm= - 2\pi \int_{r\le R} \!\!\!d^{d-1}x\  \frac{R^2-r^2}{2R}\ T_{\te\te}\ +\ c'\,.
\labell{modu}
\ee
where the constant $c'$ is fixed by demanding that the corresponding density matrix is normalized with unit trace.
We return to this specific example in a moment but first continue with our general considerations. 

In particular, given eq.~\reef{simple77}, we may now write the effective twist operator as
\be
\tsigma_n =  e^{-(n-1)\Hm}\,.
\labell{toomuch}
\ee
Hence we have produced an expression for the (effective) twist operator itself where we can easily consider the limit $n\to1$ by simply
expanding the right-hand side above in powers of $(n-1)$. Again, the purpose of this expansion is to investigate the (universal) behaviour 
of correlators involving $\sigma_n$ in the limit $n\to1$. Of course, the above expression \reef{toomuch} will only prove useful in situations
where the modular Hamiltonian is known and so the case of interest here, \ie a CFT reduced to a spherical region,
is one such situation. Hence as we will see below, we can provide evidence supporting eq.~\reef{toomuch} using the 
explicit expression for the modular Hamiltonian in eq.~\reef{modu}.

\subsection{Consistency checks}
\labell{amstel}

In the following, we provide evidence confirming eq.~\reef{toomuch} by focussing on the special situation of a spherical entangling surface in a CFT,
for which the modular Hamiltonian is given by eq.~\reef{modu}. In particular, we perform two consistency checks of eq.~\reef{toomuch} using this latter expression to evaluate the correlator
$\langle T_{\al\beta}\,\sigma_n\rangle$ and a certain contribution to $\langle \sigma_n\,\sigma_n\rangle$, both in the limit $n\to1$.
\vskip 1em

\noindent{\bf Correlator with the stress tensor:} Here, we evaluate the correlator $\langle T_{\al\beta}\,\sigma_n\rangle$ using eq.~\reef{toomuch} and then compare the result to eq.~\reef{stressfin}, when taking the limit $n\to1$. 
With eq.~\reef{toomuch}, the desired correlator becomes
\beqa
&\langle T_{\al\beta}(\te=0,x) & \sigma_n \rangle =\langle T_{\al\beta}(\te=0,x) \ \tsigma_n \rangle_1
=\left\langle T_{\alpha\beta}(\te=0,x)\ e^{-(n-1)\Hm}\right\rangle_1
\nonumber\\
&&=
 -(n-1)\, \langle T_{\alpha\beta}(\te=0,x)\, \Hm\rangle_1 + \cdots  \labell{fancy} \\
&&= 2\pi\,  (n-1) \int_{y^2\le R^2}\!\!\! d^{d-1}y \left(\frac{R^2-y^2}{2R}\right) \langle T_{\al\beta}(\te=0,x)\, T_{\te\te}(\te=0,y)\rangle_1 + \cdots  \nonumber 
\eeqa
where the ellipsis indicates terms with higher powers of $(n-1)$.
Note that we have restricted our attention to the case where both the stress tensor and the twist operator lie in the hyperplane at $\te=0$, which simplifies the subsequent
calculations somewhat. Now, substituting eq.~\reef{emt2p} into eq.~\reef{fancy} above yields,
\beqa
\langle T_{\al\beta}(\te=0,x) \ \sigma_n \rangle &=&
 2\pi\,  (n-1)\,C_T\int_{y^2\le R^2}\!\!\! d^{d-1}y \left(\frac{R^2-y^2}{2R}\right) {\delta_{\alpha t_{\mt E}}\delta_{\beta \te}-{1\over d}\delta_{\alpha\beta} \over |x-y|^{2d}}+ \cdots  \nonumber 
 \\
&=& {(n-1) \Omega_d \over (d+1)}
 {R^d\over|r^2-R^2|^d} \,C_T \, \big(\delta_{\alpha t_{\mt E}}\delta_{\beta \te}-{1\over d}\delta_{\alpha\beta} \big) + \cdots \, .
\labell{first1}
\eeqa
where $r=|x|$. The final result is written so as to accomodate both situations where $r>R$ and $r<R$ --- in the latter case, the integral in the first line must be regulated along the lines of
the discussion around eq.~\reef{delh}. Hence our final result for this correlator can be written as 
\beqa
 \langle T_{\te\te}(\te=0,x) \,\sigma_n\rangle&=&\ (n-1)\,{(d-1) \Omega_d \over d\,(d+1)}
 {R^d\over|r^2-R^2|^d} \,C_T+\cdots  \, , 
\labell{compar11}\\
 \langle T_{ij}(\te=0,x) \ \sigma_n\rangle&=&-(n-1)\,\delta_{ij}\,{ \Omega_d \over d\,(d+1)}
 {R^d\over|r^2-R^2|^d} \,C_T+\cdots \, ,
\nonumber\\
\langle T_{i \te}(\te=0,x) \,\sigma_n\rangle&=&0\,,
\nonumber
\eeqa
where we have adopted Cartesian coordinates in the $(d-1)$ directions orthogonal to $\te$. 
%

Next we turn to evaluating the same correlators using eq.~\reef{stressfin}. We will focus on reproducing the first line of eq.~\reef{compar11}
since the remaining components follow from the tracelessness of the stress tensor and the spherical symmetry of the twist operator. Using eqs.~\reef{factx} and \reef{CReq}, we find
\be
\Omega(\te=0,x)=\frac{2\,R^2}{|r^2-R^2|}\,,\qquad\quad
\frac{\del\taue}{\del\te}\Big|_{\te=0,x}=\frac{2\,R^2}{R^2-r^2}\,.
\labell{usefull8}
\ee
Substituting these expressions into the first line of eq.~\reef{last} and replacing the difference of the energy densities using eq.~\reef{heavy},
we find
\be
\langle T_{\te\te}(\te=0,x) \ \sigma_n\rangle=\frac{d-1}{2\pi\,n}
 \({2R\over|r^2-R^2|}\)^d \,h_n \, . 
\labell{almost88}
\ee
Now recall that $h_n$ vanishes in the limit $n\to1$ and hence when $n\simeq 1$, we can approximate the above expression as
\be
\langle T_{\te\te}(\te=0,x) \,\sigma_n\rangle=(n-1)\,\frac{d-1}{2\pi}
 \({2R\over|r^2-R^2|}\)^d \ \del_n h_n|_{n=1}+\cdots \, . 
\labell{almost99}
\ee
Of course, the next step is to replace $\del_n h_n|_{n=1}$ in the above expression using eq.~\reef{interest1}. For the purposes of the present
comparison, we rewrite the latter equation as
\be
\partial_n h_n|_{n=1}=\frac{2\pi\,\Omega_{d}}{2^{d}\,d\,(d+1)}\ C_T\,.
\labell{interest2}
\ee
Then substituting this expression into eq.~\reef{almost99} yields precisely the correlator given in the first line of eq.~\reef{compar11}.

Hence we have shown that to leading order in $(n-1)$, eq.~\reef{toomuch} reproduces the correct correlator $\langle T_{\al\beta}\,\sigma_n\rangle$. 
The above calculations were limited to the case where both operators lie in the hypersurface $\te=0$ but it would be straightforward to extend the comparison for
operator insertions at arbitrary relative positions. It would also be interesting to extend this comparison to higher orders in $(n-1)$, but
a comment is in order on this point. As discussed earlier in this section, in eq.~\reef{fancy}, we are inserting
the stress tensor into a single copy of the CFT. In contrast, the correlator in eq.~\reef{compar00}, and hence eq.~\reef{almost88}, involves an insertion of the full stress
tensor of all $n$ copies of the CFT. However, it is straightforward to verify that in the limit $n\to1$, differences between the two cases only arise at order $(n-1)^2$. 
Of course, for this comparison to succeed at higher orders, one must be careful to keep track of these differences.

\vskip 1em

\noindent{\bf Correlator of two twist operators:} As in the previous section, we can combine eqs.~\reef{modu} and \reef{toomuch} to produce
a fairly explicit expresssion for the effective twist operator for single spherical region.
Now one might consider whether this can be used to give useful information when we deal with multiple spherical regions, as considered in in \cite{extra1,cardy99}, but it turns out that this
is a subtle issue, as we will discuss in section \ref{discuss}. In any event, as a step in this direction, we will examine the correlator
$\langle \tsigma_{n,1}\,\tsigma_{n,2}\rangle_1$ for two spherical regions, in the following. 

To produce a tractable calculation, we look for the leading contribution to $\langle \tsigma_{n,1}\,\tsigma_{n,2}\rangle_1$ in the limit $n\to1$ and we also focus on the leading 
large-distance behaviour, \ie if the two spheres have radii $R_1$ and $R_2$ and their centers are separated by a distance $r$, then we consider $r\gg R_{1,2}$.
Further, we position both spheres in the hyperplane $\te=0$ with the first sphere centered at $\vec{x}_{c,1}$ while the second is positioned
at $\vec{x}_{c,2}$ with $|\vec{x}_{c,2}- \vec{x}_{c,1}|=r$. Adapting eq.~\reef{modu} to these positions yields
\be
H_{m,i}= - 2\pi \int_{|\vec{x}_i-\vec{x}_{c,i}|<R_i}\!\!\!\! \!\!\!\!\!\!d^{d-1}x_i\  \frac{R_i^2-|\vec{x}_i-\vec{x}_{c,i}|^2}{2R_i}\ T_{\te\te}(\te=0,\vec{x}_i)\ +\ c'\,.
\labell{modu2}
\ee
To determine the leading $n\to1$ behaviour, we replace the effective twist operators with eq.~\reef{toomuch} and
expand each of the two exponentials $\exp\[-(n-1) H_{m,i}\]$ to leading order in $(n-1)$. This yields
\bea
\langle \tsigma_{n,1}\,\tsigma_{n,2}\rangle_1 &\simeq&4\pi^2 (n-1)^2 \prod_{i=1}^2\,\int_{|\vec{x}_i-\vec{x}_{c,i}|<R_i}\!\!\!\! \!\!\!\!\!\!d^{d-1}x_i\  \frac{R_i^2-|\vec{x}_i-\vec{x}_{c,i}|^2}{2R_i}\ 
 \langle T_{\te\te}(\te=0,\vec{x}_1)\, T_{\te\te}(\te=0,\vec{x}_2)\rangle \nonumber \\
& \simeq&  4\pi^2 (n-1)^2\, \Omega_{d-2}^2\prod_{i=1}^2\,\int_{r'<R_i}\!\!\!\!dr'\,{r'}^{d-2}\   \frac{R_i^2-{r'}^{2}}{2R_i}\ {d-1\over d}\,{C_T \over r^{2d}}
\ +\ \cdots\,,
\labell{grumpy2}\\
&=& \frac{4\pi^2 \,\Omega_{d-2}^2}{d(d-1)(d+1)^2}\,(n-1)^2\,\({R_1\,R_2 \over r^{2}}\)^d\,C_T+\cdots
\nonumber
\eea
where we evaluated the correlator of the two stress tensors using eq.~\reef{dunce2}. However, since we only wish to determine the
leading behaviour at large separations, we have approximated $|\vec{x}_{2}- \vec{x}_{1}|\simeq r$ in this expression. 

This limit of the correlator can also be calculated independently using our results 
from section \ref{OPE} for the OPE expansion of the spherical twist operators. In particular, eq.~\reef{summ95} gives
the leading contribution of the stress tensor, which suggests that the desired
correlator is given by
\be
\langle \tsigma_{n,1}\,\tsigma_{n,2}\rangle_1= \gamma^2\,
R_1^d\, R_2^d \  \[(n-1)\,\partial_n h_n|_{n=1}\]^2\ \langle T_{\te\te}(\te=0,\vec{x}_{c,1})\, T_{\te\te}(\te=0,\vec{x}_{c,2})\rangle \,.
\labell{bigL2x}
\ee
Hence we see the appearance of the familiar coefficient $\partial_n h_n|_{n=1}$ here and further that the correlator controlling this interaction term 
is precisely the same as appears in eq.~\reef{grumpy2}. To verify that the numerical coefficient indeed
coincides with that in eq.~\reef{grumpy2}, first recall that $\gamma=2^{d-1}d/(\pi C_T)$ as given in eq.~\reef{summ95}. Then substituting in eqs.~\reef{interest1} and \reef{dunce2},
we find
\be
\langle \tsigma_{n,1}\,\tsigma_{n,2}\rangle_1 = \frac{2^{2d}\pi^d d(d-1)\,\Gamma(d/2)^2}{\Gamma(d+2)^2}\,(n-1)^2\,\({R_1\,R_2 \over r^{2}}\)^d\,C_T
\labell{bigL2}
\ee
and one can verify that the numerical prefactor here precisely matches that in eq.~\reef{grumpy2}.  Hence this agreement provides further evidence supporting our expression
for the effective twist operator in eq.~\reef{toomuch}. 

We note that the above agreement readily extends to the case where one of the operators is inserted away from the $\te=0$ hyperplane. Above, we saw that the same correlator of two stress tensors controls the long-distance and $n\to1$ limit in both calculations. Further we already verified that the overall coefficients match as well. Therefore we will continue to find agreement when the two effective twist operators are inserted at arbitrary positions. Interestingly, it seems this kind of correlator where the two insertions are not both at $\te=0$ is not something that one would naturally consider in evaluating the \ren entropy with $\langle \sigma_{n,1}\,\sigma_{n,2}\rangle$.

\section{Discussion} \labell{discuss}

In this paper, we investigated various properties of twist operators in higher dimensional CFT's. In particular, we made use
of the construction in \cite{renyi,circle4} where the entanglement entropy, as well as the \ren entropies, of a spherical region
in the flat space vacuum were related to the thermal entropy of the CFT on the hyperbolic background $S^1\times H^{d-1}$. This conformal mapping allows one to evaluate the scaling dimension of the twist operators $\sigma_n$ in terms of the energy
density in the thermal ensemble \cite{renyi}, as described in section \ref{twist}.  While it was originally motivatd by holographic studies of entanglement
entropy, this construction makes no reference to the AdS/CFT correspondence and in particular then, the resulting expression
for the conformal dimension \reef{heavy} applies for any CFT. Further, we might note that while the radius of the sphere in
flat space appears with an explicit factor of $R^d$ in eq.~\reef{heavy}, this is the only scale in the calculation and so the same scale
also appears implicitly in the temperature \reef{bus} and as the curvature scale of the hyperbolic space. Since the underlying field theory is a CFT, 
 the energy density in eq.~\reef{heavy} must produce a factor of $1/R^{d}$ leaving $h_n$ to be a pure number
which characterizes the conformal dimension of all twist operators in the theory.  

Of course, eq.~\reef{heavy} may not provide a very practical approach to determining $h_n$, \ie we must evaluate the energy density
of a higher dimensional CFT in a curved background with a curvature scale $R$ at temperatures of order $1/R$.  However, we were able
to use this expression to construct an expansion \reef{series} of the conformal dimension in power series around $n=1$ (where $n$ is
the order of the twist operator).\footnote{We note that this expansion was recently extended to include a chemical potential in discussing 
a new class of `charged' \ren entropies \cite{charge13}.} Further, $h_{n,k}=\partial_n^k h_n|_{n=1}$, \ie
the coefficient of the term proportional to $(n-1)^k$ in eq.~\reef{series}, 
is completely detemined by the $(k+1)$- and $k$-point correlation functions of the stress tensor in flat space. Hence, we showed that the first derivative of
the conformal dimension had a simple universal form \reef{interest1} which was fixed by $C_T$, the central charge appearing in the two-point
correlator \reef{emt2p} of the stress tensor. This univeral form was first discovered in \cite{renyi} where it was found to apply to a variety of holographic
CFT's but here, we established that it is a completely general result that applies in any higher dimensional CFT. We also showed that $\partial_n^2 h_n|_{n=1}$
has a similar universal form \reef{q2h} which can be expressed in terms of $\aha$, $\bha$, $\cha$, the three parameters which determine the three-point
function of the stress tensor \cite{Osborn0,Erdmenger0}.

In section \ref{compare} and appendix \ref{appfree}, we verified the universal expressions in eqs.~\reef{interest1} and \reef{q2h} with explicit calculations
in a variety of holographic models, as well as for a free massless fermion and for a free conformally coupled scalar. However, we must remind the reader
that for the free conformally coupled scalar in $d\ge3$, the heat kernel calculations in appendix \ref{appfree} produced a result for $h_{n,2}$ which was not in agreement with eq.~\reef{hothouse2},
\ie our general formula \reef{q2h} with the free field values for $\aha$, $\bha$ and $\cha$ substituted in.  It remains a challenge to explain this discrepancy at this point and we remind the reader that eq.~\reef{q2h} successfully 
passed  all of our holographic tests, as well as agreeing with the heat kernel computations for free massless fermions. Addressing this challenge may in fact uncover
some new perspectives on \ren entropies and twist operators. Further, we might also point out that similar discrepancies appears in applying the
approach of \cite{circle4} to evaluate the entanglement entropy of a Maxwell field in four dimensions, \eg \cite{ugly}.

In section \ref{OPE}, we considered the `operator product expansion' of spherical twist operators in higher dimensional CFT's. 
In particular, at an intermediate step, the calculation in section \ref{twist} of the scaling dimension 
produced the correlation function $\langle T_{\al\beta}\,\sigma_n\rangle$ in eq.~\reef{stressfin}. By examining this correlator in the
limit where the separation of the two operators was much larger than the radius of the sphere, we were able to evaluate the
coefficient of the stress tensor in the OPE of the twist operator, with the result given in eq.~\reef{summ95}. Again, this result applies for
general CFT's with the coefficient being determined by the ratio $h_n/C_T$. 

In principle, analogous calculations using the conformal mapping  in section \ref{twist} could be made to evaluate other terms
in the OPE of a spherical twist operator. In particular, if one could evaluate various thermal correlators in the background $S^1\times H^{d-1}$,
then they can be mapped to the corresponding correlators in the $n$-fold cover of flat space. By carefully examining the latter in the limit of
large separations, one should be able to interpret them as flat space correlators with local operators inserted at the position of the twist operator, \ie
with the OPE of the spherical twist operator. Again, the necessity to first evaluate the thermal correlators may make this an impractical approach 
for determining the OPE coefficients except in special cases. However, one observation is that generally we do not expect any local operators apart from
the stress tensor to acquire an expectation value in the thermal bath. Hence the only terms in the OPE with a single local operator in 
a single copy of the CFT would involve (normal ordered) products of the stress tensor, \ie descendents of the identity operator. However, this does not
preclude the appearance of terms involving the tensor product of operators in multiple copies of the $n$-fold replicated CFT \cite{matt99}
--- see also \cite{extra1,fuse1,quantum}. Such contributions to the OPE would be revealed in the calculation described above by thermal correlators
with several local operators suitably spaced along the thermal circle. It would be interesting to explicitly carry out such calculations in a holographic
framework or with free fields.

In eq.~\reef{simple88}, we proposed the construction of a new `effective twist operator' $\tsigma_{n}$ which acts within a single copy of the QFT
to reproduce correlators with the twist operator. We also provided 
a simple representation of this effective twist operator in terms of the modular Hamiltonian in eq.~\reef{toomuch}.
Further, a few preliminary consistency checks of eq.~\reef{toomuch} were given in section \ref{amstel}. Our arguments in section \ref{small}
are quite general and so eq.~\reef{toomuch} will apply for any quantum field theory, not just a CFT, and for any entangling
geometry, not just a spherical entangling surface. One conclusion that is drawn from eq.~\reef{toomuch} is that the reduced density matrix is 
fully determined by the twist field $\sigma_2$, \ie $\tsigma_2=\rhoa$. At first sight, this result may seem
surprising because one needs at least all the \ren entropies to get the entanglement spectrum, \eg \cite{calabrese}.
However, the \ren entropies provide a single number from the expectation value
of each twist operator and so it should be expected that  reconstructing the density matrix requires evaluating an
infinite number of such expectation values. In contrast, $\tsigma_2$ (or any single $\tsigma_n$) is an operator with which in principle, one can
calculate an infinite number of correlators. So given all of this available information, it is less surprising that one 
can reconstruct the full density matrix.

Eq.~\reef{toomuch} exhibits an apparently curious feature: On the one hand, the twist
operator is assumed to be an object which is independent of the quantum state of the underlying QFT but there will be a distinct
modular Hamiltonian for each different state on a fixed region $A$,  \ie the modular Hamiltonian is
defined with $\rhoa= \exp\[-H_m\]$ in eq.~\reef{important}.  However, the origin of this apparent disparity is straightforward.
Recall that the effective twist operator is constructed from the original twist operator, as in eq.~\reef{simple99}, by 
integrating out the ($n$--1) copies of the QFT apart from the first copy. Certainly performing this path integral will produce an operator that depends
on the quantum state since these other copies of the QFT will be in the same state as the final QFT in which $\tsigma_n$ operates.
Hence it would be interesting to test eq.~\reef{toomuch} in a situation where the region under consideration was fixed but the density matrix was changed.
One might observe a similar discrepancy in dimensionality: The twist operator is a ($d$--2)-dimensional
surface operator inserted along the entangling surface at the boundary of the region
on which the density matrix is defined. In contrast, the modular Hamiltonian is in general a nonlocal object but certainly it involves
integrals of operators over the entire region --- for example, recall eq.~\reef{modu}. However, it is again clear from the path integral construction
 \reef{simple99} of the effective twist operator that it is a nonlocal object with support across the entire region A.

Eq.~\reef{toomuch} allows us to to examine the behaviour of correlators of the twist operator with other operators in the limit $n\to1$. 
Alternatively, we can consider an expansion for small $(n-1)$ of the twist operators themselves. In particular then, the derivatives of this expression
at $n=1$ yield:
\be
\del_n \sigma_n|_{n=1} = -\Hm\,, \qquad \del^2_n \sigma_n|_{n=1} = \Hm^2 \,,
\qquad \del^k_n \sigma_n|_{n=1} = (-)^k\,\Hm^k\,.
\labell{toomuch2}
\ee
Note that we are writing these expressions for the twist operators themselves, rather than the effective twist operators. To illustrate
the sense in which these equalities hold, we consider applying the first derivative to one of the correlators\footnote{That is, we consider the correlator of the twist 
operator $\sigma_n$ with some collection of operators $\cal X$, all of which act in a single copy of the QFT or are inserted on a single sheet of the $n$-fold covering geometry.} 
discussed in section \ref{small},
\beqa
\lim_{n\to1}\del_n\langle\sigma_n\,\X\rangle&=&\lim_{n\to1}\del_n\langle\tsigma_n\,\X\rangle_1
\nonumber\\
&=&\lim_{n\to1}\del_n\langle e^{-(n-1)\Hm}\,\X\rangle_1
\nonumber\\
&=&-\langle \Hm\,\X\rangle_1
\labell{smaller}
\eeqa
Of course, this result for the first derivative is essentially equivalent to the recent result of \cite{newmisha}. Using techniques developed in \cite{vlad},
the latter argues that evaluating correlator on a manifold with an infinitesimal conical defect along a certain codimension two surface
is equivalent to the same correlator in flat space but with an extra insertion of the entanglement Hamiltonian. The correspondence
of this result with the first derivative in eq.~\reef{toomuch2} comes from the geometric approach to the replica trick, where one first
analytically continues the background geometry to non-integer $n$ \cite{callan} --- see also the discussion in \cite{cthem}.\footnote{We note
that this continuation is only straightforward for cases where there is a rotational symmetry about the entangling surfaces but that recent
progress \cite{squash} also allows one to consider infinitesimal variations of the geometry around $n=1$ for general entangling surfaces.}
It is interesting that eq.~\reef{toomuch2} suggests that higher derivatives also produce a universal effect on correlators in terms of insertions
of higher powers of the modular Hamiltonian.

To close, we return to the question of the modular Hamiltonian for regions with several simply-connected components. Of course, this
discussion is closely related to the work appearing in \cite{extra1,cardy99,extra2}. Recall that in section \ref{small}, an explicit expression
for the effective twist operator for single spherical region was constructed by combining eqs.~\reef{modu} and \reef{toomuch}.
Na\"ively, one may think this result can be used to give information about the entanglement for multiple spherical regions,
at least in the limit where the separations between the various regions are large
compared to the size of each sphere. For example, one might think that in evaluating the corresponding \ren entropy, the following 
provides a good approximation 
\be
Z_n(\textrm{$N$ spherical regions}) \simeq \big\langle \prod_{i=1}^N\ \tsigma_{n,i}\big\rangle_1
 =\big\langle \prod_{i=1}^N \exp\[-(n-1) H_{m,i}\]\big\rangle_1\,,
\labell{trial1}
\ee
where $H_{m,i}$ denotes the modular Hamiltonian \reef{modu} for the individual spherical region $i$. The basic assumption
in writing eq.~\reef{trial1} is that at large separations, the full modular Hamiltonian $H_m^{(\mt{N})}$ for the $N$ spherical regions is well approximated by the
sum of the modular Hamiltonians derived for each of the individual regions, \ie $H_m^{(\mt{N})}\simeq\sum_{i=1}^N H_{m,i}$.

Strictly speaking, this split of the modular Hamiltonian into a sum of terms for the individual regions cannot be true because it would
imply that the mutual information between these regions vanishes. However, in fact, it is not even a reasonable approximation since it misses
important leading order contributions. For example, in eq.~\reef{grumpy2}, we found the leading behaviour in the correlator $\langle
\tsigma_{n,1}\,\tsigma_{n,2}\rangle$ to two spherical regions in a general CFT decayed as $(R_1R_2/r^2)^d$. However, it has been shown that the corresponding decay for 
$\langle\sigma_{n,1}\,\sigma_{n,2}\rangle$ is given by $(R_1\,R_2/r^2)^{d-2}$ for a free massless scalar field \cite{cardy99,multi,shiba}\footnote{These
references investigate the decay in the mutual information but their results imply an analogous decay in the correlator of the corresponding twist operators.} and by
$(R_1\,R_2/r^2)^{d-1}$ for a free massless fermion \cite{multi}. These results explicitly demonstrate that in general the leading long-distance behaviour in 
correlator of two (spherical) twist operators is not controlled by the stress tensor, but rather by operators with a lower conformal dimension. In particular,
this arises if the CFT contains primary operators ${\cal O}_{\Delta}$ with dimension $\Delta\le d/2$ \cite{matt99}. These operators can appear in
the OPE of the twist operators in terms of the form $\sum_{i\ne j} {\cal O}_{\Delta,i} \otimes {\cal O}_{\Delta,j}$  where $i,j$ indicate the copy of the CFT. 
If $\Delta\le d/2$, these terms will dominate over the stress tensor in 
contributing to the long-distance decay in the correlator of the twist operators.
Since the individual operators ${\cal O}_{\Delta,i}$ appear in different copies of the CFT in these terms, these contributions
are not captured by the correlator of the effective twist operator $\langle \tsigma_{n,1}\,\tsigma_{n,2}\rangle$. The implicit assumption here is that $\langle \sigma_n\,\sigma_n\rangle$ is the standard
correlator as would appear in the evaluation of $\tr[\rhoa^n]$ for a region with two separated components. That is, both twist operators are interlacing
the same $n$ copies of the CFT. One could consider more `exotic' correlators where the two twist operators each connect $n$ copies of the CFT but only
one of these copies is common to both $\sigma_n$. In this situation, we would in fact expect that eq.~\reef{grumpy2} properly
describes the leading long-distance behaviour of the correlator. These two different situations are illustrated in figure \ref{rilke}. It would be
interesting to test these ideas by explicitly evaluating an example of the latter correlator in, \eg a free field theory.
\begin{figure}[h!]
\centering
\subfloat[]{\includegraphics[width=0.4\textwidth]{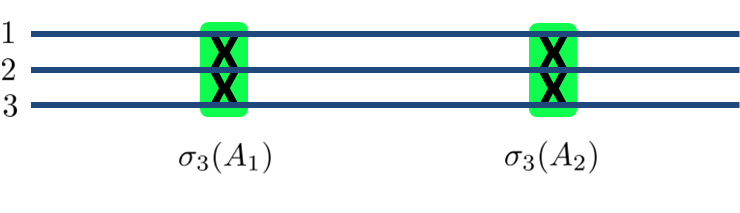}}\qquad
\subfloat[]{\includegraphics[width=0.4\textwidth]{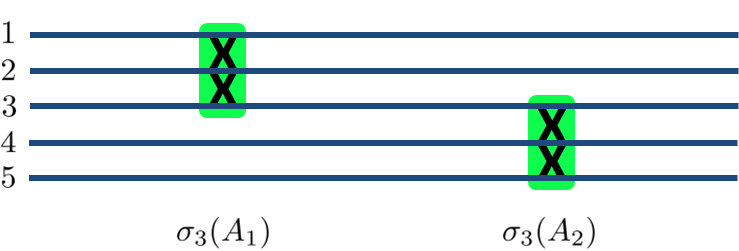}}
\caption{(Colour online)  (a) The standard three-fold geometry which would appear in the evaluation of $\tr[\rhoa^3]$ for a region with two well separated components,
$A_1$ and $A_2$. Alternatively, each component $A_i$ is delineated by a twist operator $\sigma_3(A_i)$, which connects the same three copies of the CFT. 
(b) An unconventional five-fold geometry where there is a cut through $A_1$ on sheets 1, 2 and 3 while the cut at $A_2$ runs through sheets 3, 4 and 5. 
The corresponding twist operators only have the third copy of the CFT in common. }\labell{rilke}
\end{figure}

As an extension of the above discussion, we would like to consider the appearance of so-called `teleportation'  terms in the modular 
Hamiltonian of regions with several separated components. As a specific example, the modular Hamiltonian for a massless fermion in two
dimensions for a region consisting of several disjoint intervals was constructed in \cite{nonlocal} and was observed to contain
`teleportation' terms, which connected the fermion field at points within the causal diamonds of separate
intervals. While the modular Hamiltonian is expected to be nonlocal in general, we would like to argue that this nonlocality will generically
extends to the appearance of such `teleportation' contributions. In fact, our discussion above implies that the long-distance behaviour of the
correlator $\langle \sigma_n\,\sigma_n\rangle$ is controlled by such teleportation terms when the theory contains operators with $\Delta\le d/2$.
However, let us generalize these discussions as follows: First, we observe that eq.~\reef{toomuch2}
indicates that the full modular Hamiltonian of the multicomponent region will be given by 
\be
H_m^{(\mt{N})}=-\del_n \sum_i\sigma_n(A_i)|_{n=1}
\labell{lasteq}
\ee
where $\sigma_n(A_i)$ indicates the twist operator enclosing the component $A_i$. Next, let us consider the
limit of very large separations between the different components, so that each of the individual $\sigma_n(A_i)$ can be represented by
their OPE expansion. Further as discussed above and in section \ref{OPE}, the OPE of these individual twist operators 
will typically contain terms involving operators in several different copies of the underlying QFT. However, eq.~\reef{lasteq} should only be
considered as an equality in correlators within a single copy of the QFT. That is, implicitly the last ($n$--1) copies of the QFT are trivially integrated out
on the right-hand side of eq.~\reef{lasteq} leaving an operator within the first copy. However, this implicit step of performing the path integral for the other copies of the QFT will
convert the contributions which connect multiple copies of the QFT in the individual OPE's  into teleportation terms in the full modular Hamiltonian
of the multi-component region. Of course, the appearance of such teleportation contribution is perhaps not very surprising as they simple reflect the fact that
the density matrix $\rhoa$ encodes nontrivial correlations between the different regions $A_i$. What is perhaps surprising in the two-dimensional
example considered in \cite{nonlocal} is that the modular Hamiltonian {\it is} local apart from these teleporation terms. It would be interesting to
see if this behaviour extends higher dimensional CFT's for regions including several spherical components.

\vskip 1cm

\section*{Acknowledgements} We would like to thank Horacio Casini, Eric Perlmutter and Erik Tonni
for useful comments and discussions. MS is grateful to the Perimeter Institute for hospitality and financial support during the final stages of this work.
Research at Perimeter Institute is supported by the
Government of Canada through Industry Canada and by the Province of Ontario
through the Ministry of Research \& Innovation. RCM also acknowledges support
from an NSERC Discovery grant and funding from the Canadian Institute for
Advanced Research. The work of MS is supported in part by NSF Grant PHY-1214644, and by the Berkeley Center for Theoretical Physics. The research of LYH was supported in part by the Croucher Foundation.


\appendix

\section{A useful integral}
\label{integralx}

In this Appendix, we evaluate the integral defined in eq.~\reef{integral0}, which was essential for the computation of the second derivative of the scaling dimension
 \be
 I=\prod_{i=1}^2 \[\int d\Omega_i\,\int_0^1 \! dx_i \, x_i^{d-2} (1-x_i^2)\]
 {1\over  |\vec x_1- \vec x_2|^d } ~.
 \labell{integral}
 \ee
First we note that the following relation holds within dimensional regularization
 \be
 {1\over  |\vec x_1- \vec x_2|^{d-1-2\al} } ={4^{\al}\,\pi^{(d-1)/2} \Gamma(\al)\over \Gamma \big((d-1)/2-\al\big)}\,\int {d^{d-1}p\over (2\pi)^{d-1}} {e^{i\vec p\cdot(\vec x_1- \vec x_2)} \over (p^2)^\al} \quad{\rm with}\ \   \vec x_{1,2} \in R^{d-1}~. \labell{dimreg}
 \ee
Now, we replace $1/|\vec x_1- \vec x_2|^d$ in the original expression for $I$ by the above Fourier transform with $\al=-1/2$ This substitution decouples the integrals for $i=1$ and 2 in eq.~\reef{integral} and writing $\vec p\cdot\vec x_i= p\,x_i\,\cos\theta_i$,  we can perform the polar integrals using the following identity
 \be
 \int_0^\pi de^{\pm i p x \cos\theta}\sin^{d-3}\theta d\theta=\sqrt{\pi} \({2\over px}\)^{d-3\over 2}\Gamma\Big( {d-2\over 2}\Big) J_{d-3 \over 2}(p x)~.
 \ee
As a result, we obtain
 \be
 I=- \frac{\pi^{d/2}}{\Gamma(d/2)}\,  \int {d^{d-1}p \over p^{d-4} } \[\int_0^1 dx \, x^{d-1\over 2} (1-x^2) J_{d-3 \over 2}(p x)  \]^2~.
  \ee
The integral in the square brackets can be readily evaluated 
 \be
 \int_0^1 dx \, x^{d-1\over 2} (1-x^2) J_{d-3 \over 2}(p x) = {2\over p^2} J_{d+1 \over 2}(p)~,
 \ee
while the final integral over $p$ finally yields
 \be
 I=-\frac{2^{d+3}\pi^{d-2}}{d(d+2)\,\Gamma(d-1)}   ~.
  \ee
Note that there is no contradiction between the sign of $I$ and the fact that the integrand in eq. \reef{integral} is positive definite. Indeed, the integral in eq. \reef{integral} is power law divergent, and we implicitly utilize dimensional regularization to evaluate it here. Such regularization amounts to dropping the divergent term which is positive in this case whereas the subleading correction happens to be negative.

\section{Free fields on $S^1\times H^{d-1}$}
\label{appfree}

In this appendix, we use heat kernel techniques to evaluate the scaling dimension $h_n$ in the case of massless Dirac fermion $\psi$ and conformally coupled scalar $\phi$ on the Euclidean manifold $\cM=S^1\times H^{d-1}$, where $S^1$ corresponds to Euclidean time compactified on a circle with period $\beta$. Their actions are given respectively by
 \bea
S(\phi)&=& \int_{\cM}\!\!d^dx\,\sqrt{g}\ {1\over 2}
\Big(\left(\nabla\phi\right)^2+{d-2\over 4(d-1)} \, \cR\, \phi^2\Big)~,
\labell{sact}
 \\
S(\psi)&=& \int_{\mathcal{M}}\!\!d^dx\,\sqrt{g}\  \bar\psi \slashed\nabla\psi~,
\labell{fact}
 \eea
where $\cR$ is the Ricci scalar of the background geometry and we explain our spinor conventions in what follows.

For either of the above classes of theories, the partition function is
Gaussian and can be exactly evaluated using the heat kernel approach,
\eg \cite{vass}
 \be
\log Z^{(s)}={e^{i2\pi s}\over 2} \int {dt \over
t} \text{Tr}\,K^{(s)}_{\cM}\ e^{-t m_{s}^2}~,
 \label{spartition}
 \ee
where $K^{(s)}_{\cM}(t,x,y)$ with $x,y\in\cM$ is the heat
kernel of the corresponding {\it massless} wave operator on $\cM$.
The trace of the heat kernel involves taking the limit of coincident
points, \ie $y\to x$, and integrating over the remaining position $x$.
Of course, a trace is also taken over the spinor indices in the case of
the spin-$1/2$ field --- see below. In the above expression and
throughout the following, we use $s= 0$ or ${1/2}$ to indicate the
scalar or fermion cases, respectively. Finally $m_{s}$ denotes
the `effective' mass of the field under study. For the fermion, we have
simply $m_{s={1/2}}=0$, however, given the non-minimal coupling of
the scalar, we have
 \be
 m_{s=0}^2=-{(d-2)^2\over 4 R^2}~, \label{mass0}
 \ee
where we used $\cR(H^{d-1})=-(d-2)(d-1)/R^2$ for a hyperbolic space of radius $R$.

The wave operators are separable on the product manifold
under consideration and hence the heat kernel on $\cM$ can be
expressed as the product of the two individual heat kernels on $S^1$
and $H^{d-1}$, \ie
 \be
 K^{(s)}_{\cM}=\ K_{S^1}^{(s)}\,K_{H^{d-1}}^{(s)}\,,
 \label{sepheat}
 \ee
where for brevity we have suppressed the arguments of the heat kernels
here. This separation of variables is less evident in the case of the
spin-${1\over 2}$ field due to spinor structure of the heat kernel and we justify it later on.  Given eq.~\reef{sepheat}, one can
write
 \be
\text{Tr}\,K^{(s)}_{\cM}= \text{Tr}\,K_{S^1}^{(s)}\
\text{Tr}\,K_{H^{d-1}}^{(s)}\,,
 \label{traces}
 \ee
where each trace on the right-hand side involves an integration over the
corresponding component of the product manifold. In the case of spin-${1\over
2}$ field, there is an additional trace over spinor indices. 

Finally, the partition function can be used to evaluate thermal energy density
 \be
\E(\beta)=-{1\over R^{d-1}V_{\Sigma}}{\del\over\del\beta}\log Z^{(s)}(\beta)~,
\labell{thermE}
 \ee
where, as in the main text, $R^{d-1} V_{\Sigma}$ is the (regulated) volume of $H^{d-1}$ --- see \cite{renyi}. The energy density is an essential ingredient in the computation of the scaling dimension of the spherical twist operator using eq.~\reef{heavy}.

\subsection{Conformally coupled scalar}

The heat kernel on a circle can be evaluated using the method of images. It is given by an infinite sum of  heat kernels on an infinite line shifted by an integer times the inverse temperature, \ie $n \beta$, with respect to each other. The latter is necessary to maintain periodic boundary conditions for scalar field on a circle. As a result, we get
 \be
 \int_{S^1} K_{S_1}^{(1/2)}(t,x,x)=\frac{2\beta}{\sqrt{4 \pi t}}\sum_{k=1}^{\infty} e^{-\frac{k^2 \beta^2}{4 t}} \,, 
 \labell{sheatcircle}
 \ee
where the $k=0$ term has been suppressed since it represents zero temperature limit and simply shifts the free energy by a constant.

The scalar heat kernel on the hyperbolic space can be found in a vast literature, \eg see \cite{scaheat}
 \bea
K_{H^{2\ell+1}}^{(0)}(t,x,y)&=& {1\over (4\pi t)^{1/2}} \left({-1\over 2\pi R^2
\sinh\rho}\frac{\del} {\del \rho}\right)^\ell
 e^{-{\ell^2 t\over R^2}-\frac{\rho^2 R^2}{4t}} ,
 \label{scalarheat} \\
K_{H^{2\ell+2}}^{(0)}(t,x,y)&=&  e^{-{(2\ell+1)^2\over 4R^2}t}\left({-1\over 2\pi
R^2 \sinh\rho}\frac{\del} {\del \rho}\right) ^\ell  f_{H^2}^{(0)}(\rho, t),
  \label{scalarheat2}
 \eea
where $\ell$ is $0$ or a positive integer, $\rho$ is the geodesic distance between $x$
and $y$ measured in units of $R$, and
 \be
f_{H^2}^{(0)}(\rho, t)={\sqrt{2} R\over (4\pi t)^{3/2}   } \int_\rho^\infty
{\tilde\rho \, e^{-{R^2\tilde\rho^2\over 4t} } \over
\sqrt{\cosh\tilde\rho-\cosh\rho}} \, d\tilde\rho
  ~.
 \label{2dscalarheat}
 \ee
We now turn to consider even and odd $d$ separately.

\subsubsection*{Even d}

Let us assume that $d=2\ell+2$ and take the limit of coincident points in
eq.~\reef{scalarheat}, then $K_{H^{2\ell+1}}^{(0)}(t,x,x)$ takes the following
general form \cite{cahu}
\begin{equation}
K_{H^{2\ell+1}}^{(0)}(t,x,x)=\frac{P_{\ell-1}^{(0)}(t/R^2)}{(4 \pi t)^{\ell+1/2} }
\,e^{-{\ell^2 t\over R^2}}~.
\label{oddball}
\end{equation}
From \reef{scalarheat}, it follows that for
$\ell=0$, $P_{-1}^{(0)}(x)=1$ while for $\ell>0$, $P_{\ell-1}^{(0)}(x)$ is polynomial of
degree $\ell-1$ with rational coefficients
 \be
P_{\ell-1}^{(0)}(x)=\sum_{j=0}^{\ell-1} a_{j,\ell-1}^{(0)}x^j ~.
 \label{polynomX}
 \ee
For example, the first few polynomials are given by
 \bea
 P_0^{(0)}(x)&=& 1~,
 \nonumber \\
 P_1^{(0)}(x)&=&1+{2\over 3}x~,
  \nonumber \\
 P_2^{(0)}(x)&=&1+2x+{16\over 15}x^2~,
  \nonumber \\
 P_3^{(0)}(x)&=&1+4x+{28\over 5}x^2+{96\over 35}x^3~,
   \nonumber \\
 P_4^{(0)}(x)&=&1+{20\over 3}x+{52\over 3}x^2+{1312\over 63}x^3+{1024\over 105}x^4~,
  \nonumber \\
 P_5^{(0)}(x)&=&1+10x+{124\over 3}x^2+{5560\over 63}x^3+{30656\over 315}x^4+{10240\over 231}x^5~.
 \label{spoly}
 \eea
As one may surmise from these examples, $a_{0,\ell-1}^{(0)}\equiv1$ for $\ell\ge0$.

Substituting eqs.~\reef{sheatcircle} and \reef{oddball} into eqs.~\reef{spartition} and \reef{traces}, yields
 \be
 \log Z^{(0)}(\beta)= 
 \frac{V_{\Sigma}~\beta^{1-d}}{\pi^{d/2}} \sum_{j=0}^{(d-4)/2}a_{j,\ell-1}^{(0)}\Big({\beta\over 2R}\Big)^{2j}\Gamma\Big(\frac{d}{2}-j\Big)\zeta(d-2j),
 \label{ferpartX}
\ee
Finally using eq.~\reef{thermE}, the scaling dimension \reef{heavy} takes the following form
\be
 h_n=  \frac{(2\pi)^{1-d}}{d-1}
 \sum_{j=0}^{(d-4)/2}a_{j,\ell-1}^{(0)}(2j-d+1)\pi^{2j-d/2}(n^{2j-d+1}-n)\Gamma\Big(\frac{d}{2}-j\Big)\zeta(d-2j).
 \labell{hscal}
 \ee
Differentiating this expression with respect to $n$ and comparing with eq.~\reef{delh}, yields
\be
C_T={d\over d-1}\frac{2}{\pi^{d}\Omega_{d+2}}
 \sum_{j=0}^{(d-4)/2}a_{j,\ell-1}^{(0)}(2j-d+1)(2j-d)\pi^{2j-d/2+1}\Gamma\Big(\frac{d}{2}-j\Big)\zeta(d-2j).
 \ee
However, $C_T$ is also given by eq.~\reef{CT}.
The two results agree provided that
 \be
 1=\frac{2\,\Omega_{d-1}^2}{\pi^{3d/2}\, \Omega_{d+2}}
  \sum_{j=0}^{(d-4)/2}a_{j,\ell-1}^{(0)}(2j-d+1)(2j-d)\pi^{2j+1}\Gamma\Big(\frac{d}{2}-j\Big)\zeta(d-2j).
 \ee
Using eq.~\reef{spoly} we verified that this identity holds up to $d=14$.

\subsection*{ $d=3$}

While heat kernel computation for odd $d$ is straightforward, it is much more tedeous than for even $d$ since trace of the heat kernel over even dimensional hyperbolic space cannot be represented in terms of elementary functions. Therefore, as an example, we consider $d=3$ only. Generalizations to higher odd dimensions are straightforward.   Combining eqs.~\reef{spartition}, \reef{traces}, \reef{sheatcircle} and \reef{scalarheat2}, yields
 \be
 \log Z^{(0)}(\beta)={V_{\Sigma} \beta R^3 \over 16\pi^2}\sum_{k=1}^{\infty}  \int_0^\infty d\tilde\rho \, {\tilde\rho \over \sinh{\tilde\rho\over 2}}  \int_0^{\infty} {dt\over t^3}
 \, e^{-{R^2\tilde\rho^2\over 4t}-{k^2\beta^2\over 4t}}~.
 \ee
Carrying out integration over $t$ and summation over $k$, yields
 \be
  \log Z^{(0)}(\beta)={V_{\Sigma}  \over 4\pi^2 \beta R } \int_0^\infty  {d\tilde\rho\over \sinh{\tilde\rho\over 2}}~
  {\pi^2R^2\rho^2  +\beta \sinh^2{\pi R\tilde\rho\over\beta}  \big(\pi R\,\tilde\rho \,\coth {\pi R\tilde\rho\over\beta}-2\beta\big)\over \tilde\rho^3  \sinh^2{\pi R\tilde\rho\over\beta}}~.
 \ee
Note that this integral converges. Combining now eq~.\reef{heavy} with eq.~\reef{thermE}, we obtain
 \bea
 \del_n h_n |_{n=1}&=&{2\pi^2 R^2\over V_{\Sigma}} {\del^2\over \del\beta^2}\log Z^{(1/2)}(\beta) \Big|_{\beta=2\pi R}~,
 \eea
Hence, 
 \bea
  \del_n h_n |_{n=1}&=&{\pi\over 256}~,
 \eea
Comparing this result with eq. \reef{delh} leads to $C_T={3\over 32\pi^2}$ in $d=3$. The latter agrees with  eq.~\reef{CT}.

\subsection*{$\del^2_n h_n |_{n=1}$} 
 
It is natural to use the heat kernel results to also consider the second derivative of the scaling dimension. In particular, using
eq.~\reef{hscal}, it is straightforward to evaluate $h_{n,2}$ for a conformally coupled scalar field.
When we substitute the corresponding values for $\aha$, $\bha$ and $\cha$ (see eq.~\reef{fchrg}) 
into our general formula \reef{q2h} for the second derivative, we find the expression given in eq.~\reef{hothouse2}. 
Unfortunately, it turns out that the two expressions only agree for $d=2$ and they differ in higher dimensions. We do not have a 
clear understanding of this discrepancy. We find it very challenging since eq.~\reef{q2h} successfully passed  all of our holographic tests
and further we show below that it agrees with heat kernel computations for free massless fermions.

\subsection{Dirac fermion}
We start from reviewing our spinor notation. The spinors are associated with an orthonormal frame, $e^{\mu}_a$, on  $\cM=S^1\times H^{d-1}$ satisfying
 \be
 e_{a}^{\mu}\,e_{b}^{\nu}g_{\mu\nu}=\delta_{ab}~.
 \ee
The Clifford algebra in the orthonormal frame is generated by $d$ matrices $\gamma^a$, satisfying the anticommutation relations
 \begin{equation}
 \{\gamma^a,\gamma^b\}=2\delta^{ab}~.
 \labell{clifford}
 \end{equation}
The dimension of these matrices is $2^{\lfloor{d\over 2}\rfloor}$, and the associated $d(d-1)/2$ generators of $SO(d)$ are
 \be
 \sigma^{ab}={1\over 4}[\gamma^a,\gamma^b]~.
 \ee
They satisfy the standard $SO(d)$ commutation rules
 \be
 [\sigma^{ab},\sigma^{cd}]=\delta^{bc}\sigma^{ad}-\delta^{ac}\sigma^{bd}-\delta^{bd}\sigma^{ac}+\delta^{ad}\sigma^{bc}~,
 \labell{comrul}
 \ee
and the commutator of $\sigma^{ab}$ with $\gamma^c$ is
 \be
 [\sigma^{ab},\gamma^c]=\delta^{bc}\gamma^a-\delta^{ac}\gamma^b~.
 \labell{comrul2}
 \ee

The covariant derivative of a spinor may be written in terms of $e^{\mu}_a$ as follows
 \be
 \nabla_a=e^{\mu}_a\nabla_{\mu}~, \quad
 \nabla_{\mu}=\partial_{\mu}+{1\over 2} \sigma^{bc} \omega_{\mu b c}~, \quad
 \omega_{\mu b c}=e^{\nu}_b(\del_\mu e_{c\nu}-\Gamma_{\nu\mu}^\alpha e_{c\alpha})~.
 \labell{spinorcov}
 \ee
It satisfies the following anticommutation relations \cite{DeWitt}
 \begin{equation}
 [\nabla_a,\nabla_b]\psi=-{1\over 2}R_{abcd}\sigma^{cd}\psi~.
 \labell{Dirac-comm}
 \end{equation}
In the case of free massless fermions on $\mathcal{M}$, we have
 \bea
 Z^{(1/2)}(\beta)=\det(\slashed\nabla)~,
 \eea
where we used eq.~\reef{fact} and $\slashed\nabla=\gamma^a\nabla_a$.
Since the $\gamma^{a}$ matrices are covariantly constant, one can use eqs.~\reef{clifford} and \reef{Dirac-comm} to verify the following identity
 \be
 \slashed\nabla^2=(\gamma^a\nabla_a)^2=\delta^{ab}\nabla_a\nabla_b-{R\over 4}~.
 \labell{nabla2}
 \ee
Hence, the partition function for free massless fermions can be written in the following form
 \be
 \log Z^{(1/2)}={1\over 2}\log\det (\slashed\nabla\cdot\slashed\nabla^{\dag})={1\over 2}\tr\log (-\slashed\nabla^2)~,
 \labell{ferfreengy}
 \ee
and can be evaluated using the heat kernel approach \reef{spartition} where $K^{(1/2)}_{\cM}$ is associated with operator $(-\slashed\nabla^2)$. Note that in the case of  $\cM=S^1\times H^{d-1}$, $\nabla_0=\partial_\tau$ and therefore from eq.~\reef{nabla2} $\slashed\nabla^2=\partial_\tau^2+\slashed\nabla^2\Big|_{H^{d-1}}$. As a result, one can separate the Euclidean time from the coordinates on $H^{d-1}$ and get eqs.~\reef{sepheat} and \reef{traces}.

$K_{S^1}^{(1/2)}$ can be readily evaluated. Similarly to the scalar case, it is given by an infinite sum of  heat kernels on an infinite line shifted by an integer times the inverse temperature, $n \beta$, with respect to each other and weighted by $(-1)^n$ to maintain the antiperiodic boundary conditions for the fermions on a circle
 \be
 \int_{S^1} K_{S_1}^{(1/2)}(t,x,x)=\frac{2\beta}{\sqrt{4 \pi t}}\sum_{k=1}^{\infty}(-1)^k e^{-\frac{k^2 \beta^2}{4 t}} ~\mathbb{I}_{\lfloor{d\over 2}\rfloor}\, .
 \labell{dheatcircle}
 \ee
where $\mathbb{I}_{{\lfloor{d\over 2}\rfloor}}$ is the unit matrix in $2^{{\lfloor{d\over 2}\rfloor}}$ dimensions and the $k=0$ term has been dropped from the above expression, as it corresponds to $\beta\rightarrow\infty$ (zero temperature) limit and simply shifts the free energy by a constant.


Furthermore, if $d=2\ell+2$ with $\ell=0,1,2,...$,\ie  odd dimensional hyperbolic space, then heat kernel is given by \cite{camp}
\be
 K_{H^{2\ell+1}}^{(1/2)}(t,x,y)=U(x,y) \cosh\frac{\rho}{2}  \left({-1\over 2\pi R^2}\frac{\del} {\del \cosh \rho}\right)^\ell \Big(\cosh\frac{\rho}{2}\Big)^{-1} { e^{-\frac{\rho^2 R^2}{4t}} \over (4\pi t)^{1/2}},
 \labell{diracheat}
\ee
where $\rho$ is the geodesic distance between $x$ and $y$ in units of $R$ and $U(x,y)$ is the parallel spinor propagator from  $x$ to $y$.

On the other hand, for odd $d=2\ell+3$  with $\ell=0,1,2,...$, we have \cite{camp}
\be
 K_{H^{2\ell+2}}^{(1/2)}(t,x,y)=U(x,y) \cosh\frac{\rho}{2}  \left({-1\over 2\pi R^2}\frac{\del} {\del \cosh \rho}\right)^{\ell} \Big(\cosh\frac{\rho}{2}\Big)^{-1} f_2(\rho, t),
 \labell{diracheat2}
\ee
where
 \be
  f_2(\rho, t)={R\sqrt{2} \Big(\cosh {\rho \over 2}\Big)^{-1}\over (4\pi t)^{3/2}   } \int_\rho^\infty
  {\tilde\rho \cosh {\tilde\rho \over 2} e^{-{R^2\tilde\rho^2\over 4t} } \over \sqrt{\cosh\tilde\rho-\cosh\rho}} \, d\tilde\rho
  ~.
  \labell{2dheat}
 \ee

The structure of $U(x,y)$ is not important for our needs, as we are interested in the limit of coincident points in which case $U(x,y)$  reduces to an identity matrix on the  $2^{{\lfloor{d\over 2}\rfloor}}$--dimensional spinor space. We should note here that according to \cite{camp}, the dimension of the spinor space associated with eq.~\reef{diracheat} is twice smaller and thus a modification of  eq.~\reef{diracheat} might be expected. However, the same reasoning presented  in \cite{camp} which leads to eq.~\reef{diracheat} can be equally well applied to the case considered here without necessity to introduce any changes. 

We turn now to use the above results to evaluate the scaling dimension $h_n$ in various dimensions. We consider separately even and odd $d$.

\subsection*{Even d}

It follows from  eq.~\reef{diracheat}  that for $d=2\ell+2$, $K_{H^{d-1}}(x,x,t)$ takes the following general form
\begin{equation}
K_{H^{d-1}}^{(1/2)}(t,x,x)=\frac{P_\ell^{(1/2)}(t/R^2)}{(4 \pi t)^{\ell+1/2} } \, \mathbb{I}_{\ell+1},
\labell{diracheat3}
\end{equation}
where $\mathbb{I}_{\ell+1}$ is an identity matrix on a $2^{\ell+1}$-dimensional spinor space, and $P_\ell^{(1/2)}(x)$ is a polynomial of degree $\ell$ with rational coefficients
 \be
 P_\ell(x)=\sum_{j=0}^{\ell} a_{j,\ell}^{(1/2)}x^j .
 \labell{dheathyper}
 \ee
In particular $a_{0,\ell}^{(1/2)}=1$, and
 \bea
 P_0^{(1/2)}(x)&=& 1~,
 \nonumber \\
 P_1^{(1/2)}(x)&=&1+{1\over 2}x~,
  \nonumber \\
 P_2^{(1/2)}(x)&=&1+{5\over 3}x+{3\over 4}x^2~,
  \nonumber \\
 P_3^{(1/2)}(x)&=&1+{7\over 2}x+{259\over60}x^2+{15\over8}x^3~,
 \non
 P_4^{(1/2)}(x)&=&1+6x+{141\over 10}x^2+{3229\over 210}x^3+{105\over 16}x^4~,
 \non
 P_5^{(1/2)}(x)&=&1+{55\over 6}x+{209\over 6}x^2+{17281\over 252}x^3+{117469\over 1680}x^4+{945\over 32}x^5~.
 \labell{polynom}
 \eea

Substituting eqs.~\reef{dheatcircle} and \reef{diracheat3} into eq.~\reef{traces} and then into eq.~\reef{spartition}, yields
\be
 \log Z^{(1/2)}(\beta)= V_{\Sigma}~\beta^{1-d}\,
 \frac{2^{d/2}}{\pi^{d/2}} \sum_{j=0}^{(d-2)/2}a_{j,\ell}^{(1/2)}\Big({\beta\over 2R}\Big)^{2j}(1-2^{2j+1-d})\Gamma\Big(\frac{d}{2}-j\Big)\zeta(d-2j),
 \labell{ferpart}
\ee
Using now eq.~\reef{thermE}, the scaling dimension \reef{heavy} can be evaluated as
\be
 h_n=  \frac{(2\pi)^{1-d/2}}{d-1}
 \sum_{j=0}^{(d-2)/2}a_{j,\ell}^{(1/2)}(2j-d+1)\pi^{2j-d}(n^{2j-d+1}-n)\big(1-2^{2j-d+1}\big)\Gamma\Big(\frac{d}{2}-j\Big)\zeta(d-2j).
 \labell{hferm}
 \ee
Differentiating this expression with respect to $n$ and comparing with \reef{delh}, yields
\be
C_T={d\over d-1}\frac{2^{d/2+1}}{\pi^{d/2}\Omega_{d+2}}
 \sum_{j=0}^{(d-2)/2}a_{j,\ell}^{(1/2)}(2j-d+1)(2j-d)\pi^{2j-d+1}\big(1-2^{2j-d+1}\big)\Gamma\Big(\frac{d}{2}-j\Big)\zeta(d-2j).
 \ee
On the other hand, according to eq.~\reef{CT}
 \be
 C_T={d\over 2}\, {2^{{\lfloor{d\over 2}\rfloor}}\over \Omega_{d-1}^2}~,
 \label{genCT}
 \ee
where ${\lfloor{d/ 2}\rfloor}$ is the integer part of $d/2$. Of course, when $d$ is even ${\lfloor{d/ 2}\rfloor}=d/2$. These two expressions for $C_T$ agree provided that
 \be
 1={4\over d-1}\frac{\Omega_{d-1}^2}{\pi^{d/2}\Omega_{d+2}}
 \sum_{j=0}^{(d-2)/2}a_{j,\ell}^{(1/2)}(2j-d+1)(2j-d)\pi^{2j-d+1}\big(1-2^{2j-d+1}\big)\Gamma\Big(\frac{d}{2}-j\Big)\zeta(d-2j)~.
 \ee
Using eq.~\reef{polynom}, we explicitly verified that this identity indeed holds up to $d=12$.

Let us now consider the second derivative of the scaling dimension given by eq.~\reef{hferm}
\be
 \del_n^2 h_n|_{n=1}=  \frac{(2\pi)^{1-d/2}}{d-1}
 \sum_{j=0}^{(d-2)/2}a_{j,\ell}^{(1/2)}(2j-d)(2j-d+1)^2\pi^{2j-d}\big(1-2^{2j-d+1}\big)\Gamma\Big(\frac{d}{2}-j\Big)\zeta(d-2j).
 \ee
Using eq.~\reef{polynom}, we can then evaluate the second derivative in various dimensions. Table \ref{table:hferm} summarizes final results up to $d=12$.  These
results precisely match the expected expression \reef{hothouse2} for $h_{n,2}$, which was derived by substituting the free field values for $\aha$, $\bha$ 
and $\cha$ in eq.~\reef{fchrg} into our general formula \reef{q2h}.
\begin{table}[ht]
\caption{ $\del_n^2 h_n|_{n=1}$ for femions in various dimensions.} 
\centering 
\begin{tabular}{c c c c c c c} 
\hline\hline 
d & 2 & 4 & 6 & 8 & 10 & 12 \\ [0.5ex] 
\hline 
$h_{n,2}$ & $-{1\over 6}$ & $-{13\over 180\pi}$ & $-{16\over 315\pi^2}$ & $-{59\over 1050\pi^3}$&$-{1504\over 17325 \pi^4}$ & $-{10960\over 63063 \pi^5}$\\ 
 [0.5ex] 
\hline 
\end{tabular}
\label{table:hferm} 
\end{table}

\subsection*{ d=3}

In the case of odd $d$ the computation is more sophisticated than for even $d$ since trace of the heat kernel over $H^{d-1}$ cannot be represented in terms of elementary functions as in eq.~\reef{diracheat3}. Therefore we illustrate a special case $d=3$ only. Generalizations to higher odd dimensions are straightforward.   Combining eqs.~\reef{spartition}, \reef{traces}, \reef{dheatcircle} and \reef{diracheat2}, yields
 \be
 \log Z^{(1/2)}(\beta)=-{V_{\Sigma} \beta R^3 \over 8\pi^2}\sum_{k=1}^{\infty} (-1)^k \int_0^\infty d\tilde\rho \, \tilde\rho \, \coth{\tilde\rho\over 2}  \int_0^{\infty} {dt\over t^3}
 \, e^{-{R^2\tilde\rho^2\over 4t}-{k^2\beta^2\over 4t}}~.
 \ee
Carrying out integration over $t$ and then summation over $k$ gives
 \be
  \log Z^{(1/2)}(\beta)={V_{\Sigma}  \over 2\pi^2 \beta R } \int_0^\infty d\tilde\rho \, \coth{\tilde\rho\over 2}~
  {2\beta^2  \sinh{\pi R\tilde\rho\over\beta} -\pi R\,\tilde\rho \, \big(\beta+\pi R\,\tilde\rho \,\coth {\pi R\tilde\rho\over\beta}\big)\over \tilde\rho^3  \sinh{\pi R\tilde\rho\over\beta}}~.
 \ee
Note that the integral converges. Combining eq.\reef{heavy} with eq. \reef{thermE}, yields
 \bea
 \del_n h_n |_{n=1}&=&{2\pi^2 R^2\over V_{\Sigma}} {\del^2\over \del\beta^2}\log Z^{(1/2)}(\beta) \Big|_{\beta=2\pi R}~,
 \non
  \del_n^2 h_n |_{n=1}&=& {2\pi^2 R^2\over V_{\Sigma}} \(  \beta {\del^3\over \del\beta^3} + 2 {\del^2\over \del\beta^2} \)\log Z^{(1/2)}(\beta)\Big|_{\beta=2\pi R}~.
 \eea
Hence, in our case we obtain
 \bea
  \del_n h_n |_{n=1}&=&{\pi \over 128}~,
  \non
  \del_n^2 h_n |_{n=1}&=& -{13 \pi \over 960} ~.
 \eea
These two results can be compared to eqs.~\reef{hothouse} and \reef{hothouse2} with $d=3$ and again we find perfect agreement.

\end{document}